\documentclass[%
  pre,
  amsmath,
  preprint,
  reprint,
]{revtex4-1}
\usepackage[utf8]{inputenc}
\usepackage[T1]{fontenc}
\usepackage{graphicx}
\usepackage{dcolumn} 
\usepackage{bm} 
\usepackage{lipsum}
\usepackage[bottom]{footmisc}
\usepackage[separate-uncertainty=true]{siunitx}
\usepackage{dsfont}
\usepackage{braket}
\usepackage{commath}
\usepackage{amssymb}
\usepackage{mathptmx}
\usepackage{glossaries}
\usepackage{xcolor}
\usepackage{lipsum}
\renewcommand*\vec[1]{\mathbf{\bm{#1}}}
\newcommand{\ten}[1]{\bm{#1}}
\DeclareMathOperator{\snabla}{\nabla_{\vec{n}}^{\circ}}
\DeclareMathOperator{\slaplace}{\Delta_{\vec{n}}^{\circ}}
\DeclareMathOperator{\laplace}{\Delta}
\def\I{\mathrm{i}}
\DeclareMathOperator{\Tr}{Tr}
\renewcommand{\Re}{\operatorname{Re}}
\renewcommand{\Im}{\operatorname{Im}}
\newcommand{\prT}[1]{\vec{P}_{#1}^{\perp}}
\def\onedot{\(\mathsurround0pt\ldotp{}\)}
\def\cddot{
  \mathbin{\vcenter{\baselineskip.67ex
    \hbox{\onedot}\hbox{\onedot}}%
  }}%

\def\LRA{\Leftrightarrow\mkern40mu}

\DeclareMathOperator{\intn}{\int_{\mathbb{S}^2} \mathrm{d}\vec{n}\,}

\newcommand{\sara}[1]{{\color{red} #1}}

\begin{document}
\preprint{AIP/123-QED}
\pdfinclusioncopyfonts=1
\title[Emergent active pattern formation in external fields]{Emergent pattern formation of active magnetic suspensions in an external field}

\author{Fabian R.  Koessel}

\affiliation{ Institute of Physics, Johannes Gutenberg-University, Staudingerweg 7-9,
55128 Mainz, Germany
}%

\author{Sara Jabbari-Farouji}
  \email{Correspondence to: s.jabbarifarouji@uva.nl.}
 \affiliation{%
Institute of Physics, University of Amsterdam, 1098 XH Amsterdam, The Netherlands }
 \affiliation{Institute of Physics, Johannes Gutenberg-University, Staudingerweg 7-9,
 55128 Mainz, Germany\\}

\date{\today}

\begin{abstract}
We study  collective self-organization of weakly magnetic active suspensions in a uniform external field by analyzing a mesoscopic continuum model that we have recently developed.  Our model is based on a Smoluchowski equation for a particle probability density function in an alignment field coupled to a mean-field description of the flow arising from the activity and the alignment torque.
Performing linear stability analysis of the Smoluchowski equation and the resulting orientational moment equations combined with non-linear 3D simulations,  we provide a comprehensive picture of instability patterns as a function of strengths of activity and magnetic field.  For sufficiently high activity and moderate magnetic field strengths, the competition between the activity-induced flow and external magnetic torque renders a homogeneous polar steady state  unstable. As a result, four distinct dynamical patterns of collective motion emerge. The instability patterns for pushers include traveling bands  governed by  \emph{bend-twist} instabilities and dynamical  aggregates. For pullers,  finite-sized and system spanning pillar-like concentrated regions predominated by \emph{splay} deformations emerge which migrate in the field direction. Notably, at very strong magnetic fields, we observe a reentrant hydrodynamic stability of the polar steady state.

\end{abstract}

\maketitle
-----------
\section{\label{sec:intro} INTRODUCTION}



Self-propelled systems such as birds, fire ants and bacteria exhibit fascinating patterns of collective motion.
Unraveling the physical principles governing collective self-organization of such autonomous systems  have attracted tremendous attention in recent years. The efforts to understand the collective effects in self-propelled systems have led to emergence of the interdisciplinary field of active matter, see for example ~\cite{vicsek_2012,marchetti_hydrodynamics_2013,elgeti_physics_2015}. Active matter is a fundamentally non-equilibrium class of materials  which consist of particles transforming the ambient energy to some form of mechanical motion at the individual level.
 Many studies have focused on elucidating the influence of interparticle interactions on the collective behavior of active systems. 
It is found that the interplay between self-propulsion alone with simple short-ranged interactions  in minimal models such as active Brownian particles with steric interactions~\cite{wysocki_cooperative_2014,stenhammar2014phase,speck2016} or  Vicsek model with alignment interactions~\cite{vicsek_95,vicsek_2012} leads to a rich phase behavior. Novel patterns  of collective dynamics like  dynamical clusters and traveling stripes have been identified ~\cite{speck2016,vicsek_2012} which have no counterparts in equilibrium systems.

Microswimmers, such as bacteria, algae and active colloids, belong to a special  class of active systems, which generate flows upon self-propulsion in their suspending medium. As a result,  long-ranged hydrodynamic interactions induced by the self-generated flows affect their collective behavior significantly.  
Additionally,  microswimmers display new  patterns of coordinated motion in response to external fields such as chemical gradients~\cite{adler_chemotaxis_1966,theurkauff_dynamic_2012,collectivechemo},
light~\cite{Garcia2013,martin_photofocusing_2016}, gravitational~\cite{gyrotacticmodel1,ten_hagen_gravitaxis_2014,gyro_2017,wolff_sedimentation_2013,stark_swimming_2016}
and magnetic fields~\cite{spormann_unusual_1987,guell_hydrodynamic_1988,waisbord_destabilization_2016,collectivemag}. For instance, magnetotactic bacteria driven by a sufficiently strong magnetic field migrate collectively in bands which are perpendicular to the field direction~\cite{guell_hydrodynamic_1988,spormann_unusual_1987}.

The control of collective dynamics of microswimmers via an external field offers a promising route for high-tech applications such as micro-scale \emph{cargo transport}, \emph{targeted drug delivery}, and \emph{microfluidic devices}~\cite{martel_flagellated_2009, houle_magneto-aerotactic_2016,magswim7,cargodelivery}. For instance, external magnetic field has been employed to control the rheological properties of magnetic swimmers ~\cite{vincenti_actuated_2018, alonso-matilla_microfluidic_2018}. The collective dynamics of microswimmers in an external field is nonetheless poorly understood. Specifically, the effect of interplay between long-range hydrodynamic interactions  and external fields on the pattern formation, with exception of few cases~\cite{saintillan_shear_2011,alonso-matilla_microfluidic_2018,Lushi_2012},  has been little explored.  

To make further progress in this direction, we focus on the  large-scale collective dynamics of weakly magnetic microswimmers in a uniform magnetic field. We employ the kinetic theory framework~\cite{saintillan2012kineticmf} that allows us to  overcome the size limitations of costly particle-based simulations and  to capture the large-scale patterns of active suspensions over length scales much larger than the particle size. We provide an in-depth  analysis of a kinetic continuum model that we have recently developed for dilute suspension of spherical microswimmers in an alignment field~\cite{Koessel_2019}.  Although our focus is on  weakly magnetic swimmers, the model is in principle also applicable to bottom-heavy microswimmers in a gravitational field. 

 The kinetic model couples the Smoluchowski equation for probability density function of fairly dilute active \emph{spherical} suspensions  in an alignment field to mean far field hydrodynamic interactions mainly generated by the swimmers motion.  The hydrodynamic interactions are incorporated using the leading order flow field of a force-free microswimmer that is described by a force dipole. It decays as $1/r ^2$ where $r$ is the distance from the swimmer.
 Independent of the details of motility mechanism, {\it e.g.} flagellar propulsion or  surface distortions, the majority of microswimmers can be divided according to the their far field flow into pusher (extensile) and puller (contractile) swimmers, respectively.  A pusher swimmer  uses its tail to push fluid outward along its swimming axis whereas a puller swimmer employs its front appendages to pull the fluid towards its body in the direction of swimming.  These two types of swimmers produce qualitatively different hydrodynamic flows and hence are expected to produce  distinct spatio-temporal patterns.

  We study the dynamics of both puller and pusher swimmers in a magnetic field by combining  linear stability analysis and full numerical solution of 3D non-linear  kinetic equations. Our linear stability analysis consists  of investigating the stability of the probability density function of polar steady state as well as that of its  orientational moments described by uniform density and polarization fields. Combining  the two approaches we obtain complementary insights into the nature of instabilities. At low magnetic fields, a homogeneous weakly polarized state is stable, akin to an isotropic suspension of spherical swimmers. However, for sufficiently high activity strengths and moderately strong magnetic fields, a homogeneous polar state becomes unstable for both pushers and pullers.  As we vary magnetic field and activity strengths, distinct  spatio-temporal patterns  emerge.  At moderate field  and activity strengths, pushers are driven by  bend-twist  hydrodynamic instabilities and form traveling bands perpendicular to the magnetic field.   At stronger activity and field strengths, the density-driven hydrodynamic instabilities predominate pusher suspensions leading to formation of dynamical aggregates. Pullers at moderate field and activity strengths form  system spanning pillars  parallel to the field which are  predominated by splay deformations.  However, at stronger field and activity  strengths,  they form finite-sized pillar-like concentrated regions.
 Interestingly for very strong magnetic fields a homogenous polar state becomes stable again. Hence, we observe a \emph{re-entrant hydrodynamic stability}; a hallmark of competition between alignment and hydrodynamic torques.

 The remainder of this article is organized as follows. In section \ref{sec:model}, we discuss the main ingredients of the  kinetic model for a dilute suspension of polar active particles in an alignment field. In section \ref{sec:stability},  we analyze the linear stability of homogenous polar steady state  to plane-wave perturbations  for  active polar suspensions aligned by an external field using a spectral method. Then, we calculate the stability diagram as a function of strengths of activity and magnetic field. In section \ref{sec:hd_linear_stability_of_orientational_moments}, we first derive equations of motion for the orientational moments, density, polarization  and nematic fields, using   suitable closure approximations. Then, we analyze the linear stability of  moment equations.
   In section \ref{sec:pattern}, we focus on numerical solution of the Smoluchowski equation coupled to the Stokes flow to explore the non-linear dynamics.
   We first outline our simulation method based on \emph{stochastic sampling method}. Next, we investigate the emergent spatio-temporal pattern formation varying strengths of activity and magnetic field. We particularly discuss the distinguishing features of patterns observed at different instability regimes.  Finally, our main conclusions and a discussion on comparison of linear stability analysis and non-linear dynamics solution can be found in section \ref{sec:conclusion}.

\section{KINETIC THEORY FOR ACTIVE SUSPENSIONS IN AN ALIGNMENT FIELD}%
\label{sec:model}

\subsection{Model system description}
\begin{figure}
  \centering
  \includegraphics{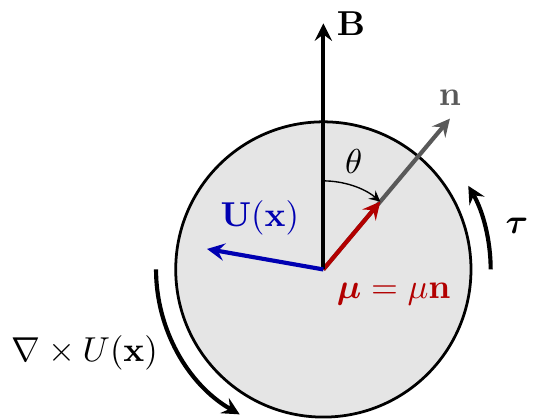}
  \caption{%
    A schematics of a spherical microswimmer with unit orientation vector \(\vec{n}\) and swimming speed \(U_0 \vec{n}\), carrying a magnetic dipole moment \(\vec{\mu}\). The dynamics of the swimmer is influenced by the local flow \(\vec{U}\) and its vorticity  \(\ten{\Omega}=  \nabla \times \vec{U}\), and an external magnetic flux density \(\vec{B}\). }  \label{fig:magnetic_microswimmer_schematics}
\end{figure}

We consider a dilute suspension of \( N \) spherical magnetic microswimmers with a hydrodynamic radius \( a \) immersed in a fluid of volume \( V \) at a number density \(\varrho_{\text{m}}=\frac{N}{V}\). We assume that the self-propulsion is generated by a force-free mechanism of hydrodynamic origin such that its far field flow, averaged over swimmer's beat cycle, is well represented by that of a point-force dipole with an effective dipolar strength \(S_{\mathrm{eff}}\)~
\cite{saintillan_theory_2015,ishikawa_suspension_2009,lauga_hydrodynamics_2009}.
\(S_{\mathrm{eff}}\) depends on the geometrical parameters of the model swimmer
~\cite{magswim3,magswim4,magswim6,adhyapak_flow_2017}, for instance on the body size \(a\) and the flagellum length \(\ell \)~\cite{adhyapak_flow_2017}.
The translational and rotational friction coefficients of the swimmer are given by \(\xi_r\) and \(\xi_t\).
Each swimmer carries a weak magnetic dipole moment \(\vec{\mu} =\mu\vec{n} \) along its body axis specified by the unit orientation vector \(\vec{n} \equiv \vec{\hat{n}}\) and has a self-propulsion velocity \(U_0\vec{n}\) as depicted schematically in Fig. \ref{fig:magnetic_microswimmer_schematics}.
The suspension is exposed to a uniform magnetic field \(\vec{B} \) that exerts an alignment torque on each swimmer. We assume that \(\mu\) is sufficiently small such that the dipole-dipole magnetic interactions at average inter-particle distance \(d_{\text{int}} \gtrsim 3 a\) are negligible relative to the thermal energy scale and no instabilities occur due to magnetic interactions.  Therefore, for volume fractions $\Phi_{\text{m}} \lesssim 0.15$ the dynamics of the system is governed by the interplay between the hydrodynamic interactions and the field-induced alignment torque.

\subsection{Conservation equation: the Smoluchowski equation}
For sufficiently low \(\varrho_{\text{m}}\), the mean-field configuration of an ensemble of swimmers at a time \(T\) can be described by a single-particle distribution function \(\Psi(\vec{X}, \vec{n}, T) \), \emph{i.e.}, the degrees of freedom of other particles have been traced out by integration. The function is normalized as
\begin{equation}
  \frac{1}{V} \int_V \mathrm{d}\vec{X} \intn \Psi(\vec{X}, \vec{n}, T) = N/V = \varrho_{\text{m}}.
\end{equation}
As such \(\Psi(\vec{X}, \vec{n}, T) / N\) describes the probability density of finding a particle with the center of mass position \(\vec{X}\) and the orientation vector \(\vec{n}\) at time $T$.
Therefore, a uniform and isotropic state can be described by the constant distribution function \(\Psi =  \varrho_{\text{m}} /4 \pi \).

The kinetic model for hydrodynamically interacting swimmers in an external field~\cite{Koessel_2019} is based on an evolution
equation for the distribution function \(\Psi( \vec{X}, \vec{n}, T) \) coupled to an equation for the mean-field fluid velocity \(\vec{U}\).
The Smoluchowski  equation for hydrodynamically interacting active particles carrying a weak magnetic dipole moment in an external field is given by
\begin{equation}
   \partial_T \Psi
    + \nabla \cdot \left[ \vec{v}_{\vec{x}} \Psi \right]
    + \snabla \cdot \left[ \vec{v}_{\vec{n}} \Psi \right]
    - \mathbb{D}\Psi
    = 0,
\end{equation}
in which \(\snabla \equiv (\mathds{1} - \vec{n} \vec{n}) \cdot \nabla_{\vec{n}}\) with dyadic product defined as $(\vec{n}\vec{n})_{ij}=n_i n_j$ denotes the angular gradient operator and \(\vec{v}_{\vec{x}}\) and \(\vec{v}_{\vec{n}}\) are the translational and rotational flux velocities resulting from a swimmer's drift. \(\mathbb{D}= D_t \nabla^2 + D_r \snabla^2 \) is the diffusion operator in which \(D_t \) and  \(D_r\) describe the effective long-time translational \(D_t\) and rotational diffusion coefficients, respectively. The diffusion coefficients can result from thermal or biological fluctuations, \emph{e.g.}, due to tumbling of bacteria in the case of rotational diffusion.
The translational flux velocity
\begin{equation}
  \vec{v}_{\vec{x}} = U_0 \vec{n} + \vec{U} \label{eq:v_t},
\end{equation}
includes the drift contributions from the self-propulsion \(U_0\vec{n}\) and an advection due to the local flow field \(\vec{U}\). 
The rotational flux velocity \(\vec{v}_{\vec{n}} \equiv \dot{\vec{n}}\) is modeled as
\begin{equation}
\label{eq:v_r}
  \vec{v}_{\vec{n}} = \prT{\vec{n}} \cdot
  \left(
    \frac{\mu}{\xi_r} \vec{B} - \ten{W} \cdot \vec{n}
  \right),
\end{equation}
in which
\(\ten{P}^\perp_{\vec{n}}=\mathds{1}-\vec{n}\vec{n} \),  describes the projection operator to the space orthogonal to the orientation vector and \(\ten{W}=\frac{1}{2} (\nabla \vec{U} - {(\nabla \vec{U})}^\top )\), with $(\nabla \vec{U})_{ij}=\partial_i U_j$,  is the vorticity tensor.
The flux velocity \(\vec{v}_{\vec{n}}\) includes the rotational drift contributions resulting from the torque due to the magnetic field and vorticity of the local flow. Using the relation between the angular velocity $\vec{\omega}$ and rate of change of orientaitona vector \(\dot{\vec{n}} = \vec{\omega} \times \vec{n}\), the first term
\(\ten{P}^\perp_{\vec{n}} \cdot \frac{\mu}{\xi_r} \vec{B}\) on the right-hand side is obtained from  the balance between the magnetic torque \(\vec{\tau} = \mu \vec{n}\times \vec{B}\) and the frictional hydrodynamic torque \(- \xi_r \vec{\omega} \)  in the overdamped and low Reynold's number limits.  The second term models the interaction of a \emph{spherical} swimmer with the local flow vorticity based on the the second Faxen's law~\cite{dhont_introduction_1996}.

From the distribution function, we define the local density field \(\rho(\vec{X}, T)\), polarization field \(\vec{p}(\vec{X}, T)\), and the nematic order parameter field \(\ten{Q}(\vec{X}, T)\), as the symmetric and traceless  parts of the zeroth, first, and second order orientational moments of \(\Psi(\vec{X}, \vec{n}, T)\) with respect to \(\vec{n}\), respectively,
\begin{align}
  \rho(\vec{X}, T) &= {\left< 1 \right>}_{\vec{n}}
    = \intn \, \Psi \; 1 \label{eq:mom_density}\\
  \vec{p}(\vec{X}, T) &
    = {\left< \vec{n} \right>}_{\vec{n}}
    = \intn \, \Psi \; \vec{n} \label{eq:mom_polarization}\\
  \ten{Q}(\vec{X}, T) & = {\left< \vec{n}\vec{n} - \mathds{1} / 3 \right>}_{\vec{n}} \nonumber\\
    &= \intn \, \Psi \; \left(\vec{n}\vec{n} - \frac{1}{3}\mathds{1}\right). \label{eq:mom_nematic}
\end{align}
These moments will be used throughout the paper in the following sections.

\subsection{Mean-field flow}
The flow field \( \vec{U}\) in Eqs.~\eqref{eq:v_t} and~\eqref{eq:v_r} may result from an
imposed external flow or from hydrodynamic interactions. In this work, we consider the case that there is no external flow and \( \vec{U}\) solely represents the self-generated flow due to motion of swimmers.
In the limit of vanishing Reynolds number, applicable to microswimmers, the fluid reacts in good approximation instantaneously to changes in the particle configuration. The mean-field flow \( \vec{U}[\Psi] \) resulting from the hydrodynamic interactions between the swimmers is well captured by the incompressible \emph{Stokes} equation
\begin{align}
  \eta \nabla^2 \vec{U} - \nabla P &+\nabla \cdot \ten{\Sigma}[\Psi] =0\\
  \nabla \cdot \vec{U} &=0,
  \label{eq:stokesflow}
\end{align}
in which \(P\)  and $\eta$ denote the isotropic pressure and  the viscosity of the suspending fluid and \( \nabla \cdot \ten{\Sigma} = \partial_{X_i} \Sigma_{ij} \hat{\vec{e}}_j\). The mean-field stress \(\ten{\Sigma}[\Psi]\) depends on the instantaneous suspension configuration encoded by \(\Psi \). In the case of microswimmers, it can be decomposed into the sum of several contributions, arising from the self-propulsion, Brownian rotations, resistance to stretching and compression by the local flow field and steric and magnetic torques. For dilute suspensions of spherical microswimmers, we neglect stresses arising from Brownian rotations (can be incorporated into active stress by modifying the prefactor), inextensibility of the particles 
 and steric torques because of their small contributions.
We only consider active stress \(\ten{\Sigma}_{\text{a}}[\Psi] \), generated by the self-propulsion of swimmers,~\cite{ishikawa_suspension_2009,lauga_hydrodynamics_2009} and a magnetic stress \( \ten{\Sigma}_{\text{m}}[\Psi] \), caused by reorientation of swimmers in the external field. Hence, the stress in our model is given by \(\Sigma[\Psi] = \Sigma_{\text{a}}[\Psi] + \Sigma_{\text{m}}[\Psi]\).

In a dilute suspension, for which the average the ratio of inter-particle distance to the swimmer  size is large, the active stress of a force-free microswimmer \(\ten{\Sigma}_{\text{a}}[\Psi] \) can be modeled as that of a point-force dipole -- the leading order non-zero singularity of the Stokes
flow~\cite{chwang_hydromechanics_1975,batchelor_stress_1970,lauga_hydrodynamics_2009}.
The active stress of a suspension of dipolar microswimmers is proportional to the nematic order tensor field  ~\cite{doi_theory_2009,saintillan2012kineticmf} as defined by Eq.~\eqref{eq:mom_nematic}:
\begin{equation}
 \ten{\Sigma}_{\text{a}}(\vec{X},T) =\Sigma_ {\text{a}}\ten{Q}(\vec{X}, T).
 \label{eq:active_stress}
  \end{equation}
It can be interpreted as a superposition of stress contributions of  all possible swimmer orientations at position \(\vec{X}\).
The strength of the active stress is determined by  the   amplitude  \(\Sigma_{\text{a}}= - \varrho_{\text{m}} S_{\mathrm{eff}}\). The sign of \(\Sigma_{\text{a}}\) determines the nature of the swimmers, being a puller \(\Sigma_{\text{a}} > 0\) or a pusher \(\Sigma_{\text{a}} < 0\).


The torque due to external field \(\vec{M}_B=\mu \vec{n} \times \vec{B}\) leads to rotation of swimmers that in turn exerts a rotational stress on the fluid while dragging the surrounding fluid layers. This results in an antisymmetric stress contribution of the form
\begin{align}
  \ten{\Sigma}_{\text{m}}
  &= {\left< \frac{1}{2} \ten{\varepsilon} \cdot \varrho_{\text{m}} \vec{M}_B(\vec{n}) \right>}_{\vec{n}} \nonumber \\
  &= {\left< \frac{ \varrho_{\text{m}} \mu B}{2} \left( \vec{n} \hat{\vec{B}} - \hat{\vec{B}} \vec{n} \right) \right>}_{\vec{n}} \nonumber \\
  &= \frac{\Sigma_{\text{m}}}{2} \left( \vec{p} \hat{\vec{B}} - \hat{\vec{B}} \vec{p} \right),
\end{align}
in which \( \vec{\hat{B}} = \vec{B} / B \), \(\Sigma_{\text{m}} \equiv \varrho_{\text{m}} \mu B\), and \(\ten{\varepsilon}\) is the Levi-Cevita symbol.
This stress contribution is identical to that of passive magnetic suspensions.  Note that the symmetric part of the magnetic stress is zero for spherical particles~\cite{ilg_magnetization_2002,ilg_magnetoviscosity_2002}. 

\subsection{Non-dimensionalization}
To facilitate the analysis of the model, we render the equations dimensionless, using the following characteristic velocity, length, and time scales: \(u_c = U_0\), \(t_c = 1/D_r\) and \(x_c=U_0/D_r\). 
Note that our choice of characteristic time and length scales are different from our previous work~\cite{Koessel_2019}.
 We rescale distribution function with the number density such that \(\psi(\vec{x}, \vec{n}, t) \equiv \Psi(\vec{x}x_c, \vec{n}, t t_c) /\varrho_{\text{m}}\), is dimensionless and \(\psi/v\) represents a probability density normalized to unity:
  \begin{equation}
  \frac{1}{v} \int_v \mathrm{d}\vec{x} \intn \psi(\vec{x}, \vec{n}, t) = 1.
\end{equation}
where \(v=V/x_c^3\).
 The form of Smoluchowski equation for \(\psi(\vec{x}, \vec{n}, t)\) remains unchanged
\begin{equation}
  \partial_t \psi
   + \nabla \cdot \left[ \vec{v}_{\vec{x}} \psi \right]
   + \snabla \cdot \left[ \vec{v}_{\vec{n}} \psi \right]
   - \mathbb{D}\psi
   = 0,
  \label{eq:hd_smoluchowski}
\end{equation}
where the gradient operator \(\nabla \equiv \partial/ \partial_{x_i} \vec{\hat{e}}_i\) is now with respect to the reduced coordinates. The dimensionless spatial and rotational flux-velocities reduce to
\begin{align}
  \vec{v}_{\vec{x}} &= \vec{n} + \vec{u} \label{eq:hd_smol_trans_flux} \\
  \vec{v}_{\vec{n}} &= \prT{\vec{n}} \cdot
  \left(
      \alpha_{\text{e}} \hat{\vec{B}} - \ten{W} \cdot \vec{n}
  \right),
  \label{eq:hd_smol_rot_flux}
\end{align}
in which \(\alpha_{\text{e}}=\frac{\mu  B}{\xi_r D_r}\) defines the alignment parameter.
Likewise, the dimensionless diffusion operator simplifies to
\begin{equation}
  \mathbb{D} = d_t \laplace + \slaplace.
  \label{eq:hd_diffoperator}
\end{equation}
where \(d_t=D_t D_r/U_0^2\) is the reduced translational diffusion coefficient.
The equation for the flow-field transforms into
\begin{align}
  \laplace \vec{u} - \nabla p &+\nabla \cdot \ten{\sigma}[\psi]=0 \label{eq:hd_stokes} \\
  \nabla \cdot \vec{u} &=0, \nonumber
\end{align}
with the dimensionless stress tensor \(\ten{\sigma}\) given by
\begin{equation}
  \ten{\sigma}=\frac{\ten{\Sigma}}{D_r \eta}  = \frac{1}{D_r \eta} \left( \ten{\Sigma}_ {\text{a}}+ \ten{\Sigma_{\text{m}}} \right).
  \label{eq:mean_stress}
\end{equation}
As such, two additional independent dimensionless parameters, the active stress amplitude \(\sigma_{\text{a}}=\frac{\Sigma_{\text{a}}}{D_r \eta} \) and the external field-induced stress amplitude \(\sigma_{\text{m}} =\frac{\Sigma_{\text{m}}}{D_r \eta} \)  appear in our model.

\section{LINEAR STABILITY ANALYSIS OF HOMOGENEOUS POLAR STEADY STATE}%
\label{sec:stability}

The set of equations~\eqref{eq:hd_smoluchowski} and~\eqref{eq:hd_stokes} forms a closed system that can be solved for the
evolution of the distribution function \(\psi \) and the flow field \(\vec{u}\) in the suspension. However, it is not presently feasible to solve these coupled equations analytically. Therefore, we resort to the linear stability analysis that provides us with some degree of predictive insight into the dynamics of the equations with respect to a suitable base state. This kind of analysis allows us to divide the parameter space into a stable region described by the base state and an unstable region with yet unknown dynamics departing from the base state. Furthermore, the linear stability analysis offers some valuable insight into the dynamics at the onset of instability.


 \subsection{Homogeneous and steady solution as a base state}%
\label{sec:hd_steady_state}

\begin{figure}
  \centering
  \includegraphics[width=0.95\linewidth]{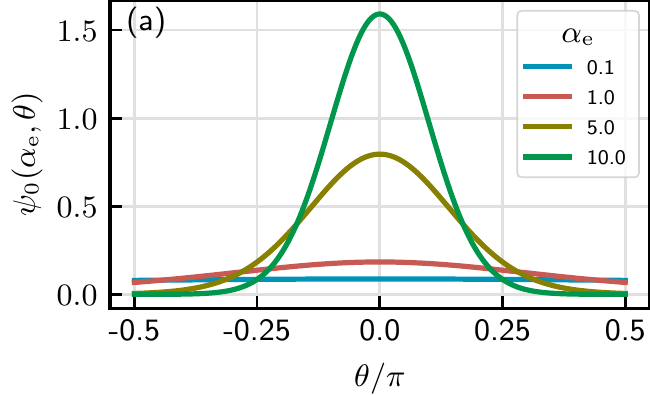}
    \includegraphics[width=0.95\linewidth]{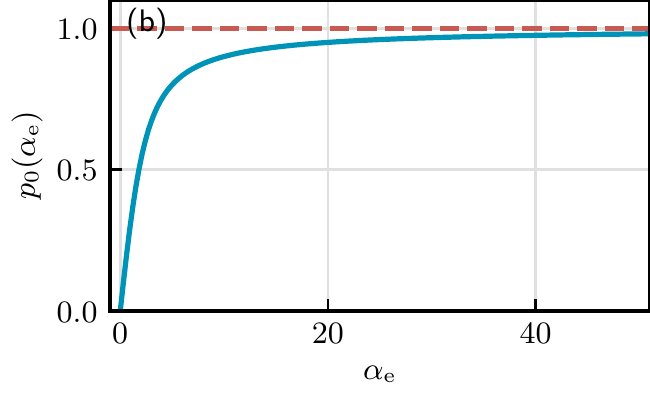}
  \caption{%
    (a) Angular distribution of the homogeneous steady state \(\psi_0(\theta)\) for different values of the external alignment parameter
    \(\alpha_{\mathrm{e}} =\frac{\mu  B}{\xi_r D_r}\).
    (b) The total polarization \(p_0\) of the steady state \(\psi_0\) as a function of 
    \(\alpha_{\mathrm{e}} \propto B\).    
  }%
  \label{fig:hd_steady_state}
\end{figure}
The external field breaks the rotational symmetry of the system but preserves translational invariance.
Thus, we first seek for spatially-uniform \( \partial_{\vec{x}} \psi_0 = 0\) and steady \( \partial_t \psi_0 = 0\) solutions of
the Smoluchowski equation~\eqref{eq:hd_smoluchowski}. Solutions of the form \( \psi_0(\alpha_{\text{e}}, \vec{n})\) will serve as base states for the linear stability analysis. For a homogenous steady state, all spatial and time derivatives in Eq.~\eqref{eq:hd_smoluchowski} vanish. The same holds for the flow field as \(\nabla \cdot \ten{\sigma} = 0 \Rightarrow \vec{u}_0 = 0\). Only the rotational flux velocity terms remain. Hence, \( \psi_0(\alpha_{\text{e}}, \vec{n})\) can be obtained by setting the total rotational flux velocity including both drift and diffusive contributions to zero, \emph{i.\ e.},
\begin{equation}
\alpha_{\text{e}}  \prT{\vec{n}} \cdot  \hat{\vec{B}} + \snabla \ln \psi_0 =0.
\end{equation}
Solving this equation yields
\begin{align}
    \psi_0(\alpha_{\text{e}}, \vec{n})
    &= \frac{\alpha_{\mathrm{e}}}{4\pi \sinh \alpha_{\mathrm{e}}}
        e^{ \alpha_{\mathrm{e}} \vec{n} \cdot \hat{\vec{B}} }
    \label{eq:hd_steady_state}
\end{align}
with the normalization \(\intn \psi_0(\alpha_{\text{e}}, \vec{n}) = 1\). This steady-state is identical to that of passive magnetic dipoles in an external field~\cite{ilg_magnetization_2002}.
For passive systems at the thermal equilibrium,  the Einstein-Stokes-Debye relation \(D_r=\frac{k_B T}{\xi_r}\) ~\cite{dhont_introduction_1996} holds and the alignment parameter  becomes   \(\alpha_{\mathrm{e}}=\frac{\mu B}{k_B T}\) which is equal to ratio between magnetic and thermal energy scales. More generally, it describes the ratio between two characteristic reorientation time scales \(\alpha_{\mathrm{e}}=\tau_{\mathrm{e}}/ \tau_{\mathrm{r}}\).
The time \(\tau_{\mathrm{r}} = \frac{1}{D_r}\) represents the average decorrelation time of a diffusive particle from its initial orientation and \(\tau_{\mathrm{e}}=\frac{\xi_r}{m B}\) is a measure of the typical alignment time of a non-diffusive dipole with the external field. The competition between alignment (order) and the randomization of orientation (disorder) determines the degree of alignment  quantified by the mean polarization \(p_0\).
It is the given by the magnitude of the polarization vector $\vec{p}_0 $:
\begin{align}
    \vec{p}_0 &= \intn \vec{n} \, \psi_0(\vec{n}) = p_0 \hat{\vec{B}} \nonumber \\
    p_0(\alpha_{\mathrm{e}}) &= \coth \alpha_{\mathrm{e}} - \frac{1}{\alpha_{\mathrm{e}}}.
    \label{eq:hd_steadystate_polarization}
\end{align}
The function \(p_0(\alpha_{\mathrm{e}})\) is identical to the well-known Langevin function appearing in the context of paramagnetism or force-extension relation of a freely jointed chain~\cite{Rubinstein}.

Assuming a magnetic field parallel to the \(z\)-axis, \emph{ i.e.} \(\vec{B}= B \hat{\vec{z}}\), without loss of generality, the homogeneous polar state with axial symmetry takes the simple form of \( \psi_0(\alpha_{\text{e}}, \vec{n})=\psi_0(\alpha_{\text{e}}, \theta)\), where \(\theta \) denotes the angle between the orientation vector and the magnetic field and it coincides with the polar angle in spherical coordinates for the orientation \(\vec{n}(\theta, \phi)=(\sin \theta \cos \phi, \sin \theta \sin \phi, \cos \phi)\). The angular dependency of the homogeneous polar steady state for different values of \(\alpha_{\mathrm{e}}\) is shown in Fig.~\ref{fig:hd_steady_state} (a). A strong external magnetic field (large \(\alpha_{\mathrm{e}}\)) results in a focused angular distribution around the magnetic field axis corresponding to \(\theta = 0\) and thus a large mean polarization \(p_0\). The functional dependency  of polarization magnitude on the alignment parameter \(\alpha_{\mathrm{e}} \) is plotted in Fig.~\ref{fig:hd_steady_state} (b). The mean polarization continuously increases with increasing \(\alpha_{\mathrm{e}} \propto B\). It asymptotically approaches a perfectly aligned state with \(p_0 = 1\) in the limit of very large \(\alpha_{\mathrm{e}}\) described by \(\lim_{\alpha_{\mathrm{e}} \to \infty} \psi_0 = \delta(\vec{n} - \hat{\vec{B}})\). In the other extreme of very low magnetic field strengths, fluctuations will increasingly decorrelate the orientation of a swimmer, leading to a flat profile in the angular distribution, \emph{i. e.}, \(\lim_{\alpha_{\mathrm{e}} \to 0} \psi_0 = 1/4\pi \), which corresponds to an isotropic suspension with \( p_0=0\).

\subsection{Linearized equations and eigenvalue problem}%
\label{sec:hd_linstab}

\subsubsection{Linear perturbation of the base state}%
\label{sec:hd_linear_stability_analysis}
We now proceed to analyze the linear stability of the homogeneous polar steady state presented in Sec.~\ref{sec:hd_steady_state}. We consider a small disturbance of the distribution function \(\psi \) with respect to \(\psi_0\).
\begin{equation*}
  \psi = \psi_0(\vec{n}) + \varepsilon \psi_p(\vec{x},\vec{n},t)
\end{equation*}
where \(| \varepsilon| \ll 1\) and \(|\psi_p|\sim \mathcal{O}(1)\). Likewise, the flow-field of the suspending medium is perturbed by,
\(\vec{u} = \vec{u}_0 + \varepsilon \vec{u}_{p}\), in which \(\vec{u}_p\) is the flow-field caused by the perurbation \(\psi_p\). The corresponding flow field of the steady state is \(\vec{u}_0 = 0\), because all the spatial derivatives on the right-hand side of Stokes equation~\eqref{eq:hd_stokes} vanish for \(\psi_0(\vec{n})\).

After neglecting terms of \(\mathcal{O}(\varepsilon^2)\) in the governing equations, we obtain the following linearized evolution equation
for \( \psi_p\)
\begin{alignat}{2}
  \partial_t \psi_{p} && =
    &- \vec{n} \cdot \nabla \psi_{p} \nonumber \\
    &&&+ 2 \vec{n} \cdot \hat{\vec{B}} \psi_{p}
       - (\prT{\vec{n}} \cdot \hat{\vec{B}}) \cdot \snabla \psi_{p} \nonumber \\
    &&&+ (\prT{\vec{n}} \cdot \ten{W}[\vec{u}_{p}] \cdot \vec{n}) \cdot \snabla \psi_0
       - 3 \psi_0 \, \vec{n} \vec{n} \cddot \ten{W}[\vec{u}_{p}] \nonumber \\
    &&&+ \mathbb{D} \psi_{p},
    \label{eq:hd_linearized_perturbed_smoluchowski}
\end{alignat}
where the double contraction $\cddot$ is defined as $(\vec{a} \vec{b} \cddot \ten{C})=a_i b_j C_{ij}$.
In our derivation, we have used antisymmetric property and tracelessness of the vorticity tensor \(\ten{W}\) and the following identities:
\begin{align*}
  \prT{\vec{n}} \cdot (\ten{A} \cdot \vec{n}) &= \Tr \ten{A} - 3 \vec{n} \vec{n} \cddot \ten{A} \\
  \snabla \cdot (\prT{\vec{n}} \cdot \vec{a}) &= -2 \vec{n} \cdot \vec{a},
\end{align*}
 which hold for any arbitrary tensor \(\ten{A}\) and vector \(\vec{a}\).
The flow field resulting from the perturbation \(\vec{u}_p\) satisfies the same momentum equation as \(\vec{u}\), but forced by the linearized
stress tensor given by
\begin{equation}
   \ten{\sigma}_p(\vec{x}, t)= \sigma_{\text{a}} {\ten{Q}}[\psi_p]+ \sigma_{\text{m}} (\vec{p}[\psi_p]\hat{\vec{B}}-\hat{\vec{B}}\vec{p}[\psi_p]])/2,
   \label{eq:stress_per}
 \end{equation}
where the time-dependence of the stress tensor stems from that of $\psi_p (\vec{x}, \vec{n},t)$.

To progress further, we Fourier-transform the linearized Smoluchowski equation, Eq.~\eqref{eq:hd_linearized_perturbed_smoluchowski} where  the Fourier transform of $\psi_p$ is defined as $\psi_p^{\mathcal{F}}= \int {d\vec{x} \psi_p e^{\I \vec{k} \cdot \vec{x} }}$ and we use the factorization ansatz $\psi_p^{\mathcal{F}}(\vec{k},\vec{n},t)=\tilde{\psi}(\vec{k}, \vec{n}) e^{\lambda(\vec{k}) t}$.  This ansatz decomposes the contribution of Fourier mode of the perturbation $\psi_p^{\mathcal{F}}$ to a time-independent amplitude $\tilde{\psi}(\vec{k}, \vec{n})$, also known as  mode shape, and an exponential growth factor with a complex growth rate given by $\lambda(\vec{k})$.
This ansatz, which arises from linearity of perturbation equations, implies that the Fourier transform of the stress tensor due to perturbation given by Eq.~\eqref{eq:stress_per}
can be written as $ \ten{\sigma}_p^{\mathcal{F}}=\tilde{\ten{\sigma}}_p(\vec{k})e^{\lambda(\vec{k}) t}$, where
 \begin{equation}
 \tilde {\ten{\sigma}}_p[\tilde{\psi}](\vec{k})= \sigma_{\text{a}}  \ten{Q}[\tilde{\psi}]+ \sigma_{\text{m}} ( \vec{p}[\tilde{\psi}]\hat{\vec{B}}-\hat{\vec{B}} \vec{p} [\tilde{\psi}])/2.
 \end{equation}
Consequently, the Fourier transform of the flow field $\vec{u}_p$ can be obtained as
\begin{align}
  \vec{u} _p^{\mathcal{F}}(\vec{k})&= \mathbb{O}^{\mathcal{F}}\cdot  (\I \vec{k} \cdot \ten{\sigma}_p^{\mathcal{F}})  
  =  \I (\mathbb{O}^{\mathcal{F}}\cdot \tilde {\ten{\sigma}}_p  \cdot \vec{k} )e^{\lambda(\vec{k}) t}
    \label{eq:hd_fourier_flow}
\end{align}
in which  \(  \mathbb{O}^{\mathcal{F}} = \frac{1}{k^2} (\mathds{1} - \hat{\vec{k}} \hat{\vec{k}})\)
 is the Fourier transform of  the Oseen tensor and \( \hat{\vec{k}} = k^{-1} \vec{k}\) is the normalized wavevector.
From above, we can see that the Fourier transform of the flow field can also be decomposed  as $ \vec{u} _p^{\mathcal{F}}(\vec{k})=  \tilde {\vec{u}}(\vec{k}) e^{\lambda(\vec{k}) t}$ where its amplitude is  explicitly given by
  \begin{equation}
  \tilde {\vec{u}}[\tilde{\psi}](\vec{k})
    = \frac{\I}{k} (\mathds{1} -\hat{\vec{k}} \hat{\vec{k}})
      \cdot \tilde {\ten{\sigma}}_p[\tilde{\psi}] \cdot \hat{\vec{k}},
    \label{eq:hd_fourier_flow}
\end{equation}

After some algebraic manipulation, the governing equation for  \(\tilde{\psi}(\vec{k}, \vec{n})\) transforms into an eigenvalue problem of the form
\begin{equation}
  \mathbb{L}[\tilde{\psi}] = \lambda \tilde{\psi},
  \label{eq:hd_eigenvalue_problem}
\end{equation}
 in which \(\mathbb{L}\) represents a linear differentio-integro-operator and
  \(\tilde{\psi}(\vec{k}, \vec{n})\) is the associated eigenvector  encoding the form of the orientational perturbation for a given \(\vec{k}\). The explicit form of  \(\mathbb{L}\) is given by
    \begin{alignat}{2}
  \mathbb{L}[\tilde{\psi}] && =
    &- \I \vec{n} \cdot \vec{k} \tilde{\psi} \nonumber \\
    &&&+ 2 \vec{n} \cdot \hat{\vec{B}} \tilde{\psi}
       - (\prT{\vec{n}} \cdot \hat{\vec{B}}) \cdot \snabla \tilde{\psi} \nonumber \\
    &&&+ (\prT{\vec{n}} \cdot \tilde {\ten{W}}[\tilde {\vec{u}}[\tilde{\psi}]] \cdot \vec{n}) \cdot \snabla \psi_0
       - 3 \psi_0 \, \vec{n} \vec{n} \cddot \tilde {\ten{W}}[\tilde {\vec{u}}[\tilde \psi]] \nonumber \\
    &&&+ \slaplace \tilde{\psi} - d_t k^2 \tilde{\psi},
    \label{eq:hd_perturbation_operator}
\end{alignat}
in which
\(
  \tilde {\ten{W}}[\tilde {\vec{u}}]
    = \frac{\I}{2}(\vec{k} \tilde {\vec{u}} - \tilde {\vec{u}}\vec{k} )
\).
Based on the form of Eq.~\eqref{eq:hd_perturbation_operator}, we note that the stability is governed by four dimensionless parameters \(d_t\), \(\alpha_{\text{e}}\), \(\sigma_{\text{a}}\) and \(\sigma_{\text{e}}\).
To determine the growth rate \(\lambda (\vec{k}) \) and hence stability of the active suspension in external field for a given set of the parameters, we need to solve the eigenvalue problem defined by equation~\eqref{eq:hd_eigenvalue_problem}. We discuss our methodology for this problem in the following subsection.

\subsubsection{Spectral method for solving the eigenvalue problem}%
\label{sec:hd_eigval}
The above analysis  shows that it is sufficient to consider plane wave perturbations of the form: \( \psi_p (\vec{x}, \vec{n}, t) = \tilde{\psi}(\vec{k}, \vec{n}) e^{\I \vec{k} \cdot \vec{x} + \lambda(\vec{k}) t} \)
and \( \vec{u}_{p}(\vec{x}, t) =\tilde {\vec{u}}(\vec{k}) e^{\I \vec{k} \cdot \vec{x} + \lambda(\vec{k}) t}\) to investigate the linear stability of the steady state.
Here,  $\Re  \lambda  $ determines growth rate and  $\Im  \lambda $ gives the frequency of a travelling wave with wavevector \(\vec{k}\).
To solve the eigenvalue problem of~\eqref{eq:hd_eigenvalue_problem}, we employ a spectral method
where we also expand the orientational dependency of the eigenfunction \(\tilde{\psi}(\vec{k}, \vec{n})\) as well as \(\psi_0\) in the basis of spherical harmonics.
We choose a spherical coordinate system in which \(\vec{B}\) is aligned with the polar axis. Denoting the polar and azimuthal angles  by \(\theta \in [0, \pi ]\) and \(\phi \in [0, 2 \pi )\), respectively, we have
\begin{equation}
  \vec{n} = (\sin \theta \cos \phi, \sin \theta \sin \phi, \cos \theta)
\end{equation}
In this coordinate system, the spherical harmonic function of degree \(l\) and order \(m = -l, \ldots, l\) is defined as
\begin{equation}
  Y_l^m (\vec{n}) =\sqrt{\frac{(2l + 1) (l+m)!}{4 \pi (l-m)!}} P_l^m (\cos \theta ) \exp(\I m \phi)
\end{equation}
where \(P_l^m (\cos \theta )\) is the associated Legendre polynomial. The spherical harmonics satisfy the
orthogonality condition:
\begin{equation}
\label{eq:ortho}
 \braket{ Y_l^m | Y_{l'}^{m'}}= \delta_{l l'} \delta_{m m'}
\end{equation}
where the scalar product is defined by
\begin{equation*}
  \braket{f | g} = \intn {f}^{*}(\vec{n}) \, g(\vec{n}),
\end{equation*}
with the star operator  \(\bullet^{*}\) representing the complex conjugatation. These functions form a complete basis on the unit sphere,
  on which we expand the mode shape \(\tilde{\psi}(\vec{k}, \vec{n})\) as
\begin{alignat}{2}
  &&\tilde{\psi}(\vec{k}, \vec{n}, t)
  &= \sum_{l=0}^{\infty} \sum_{m=-l}^{l}
    Y_l^m(\vec{n}) \, \psi_l^m(\vec{k}, t) \label{eq:th_psi_sh_expansion}\\
  \LRA &&
  \ket{\tilde{\psi}}
  &= \sum_{l=0}^{\infty} \sum_{m=-l}^{l}
    \ket{Y_l^m} \braket{Y_l^m | \tilde{\psi}},
    \label{eq:harmonic_expand}
  \end{alignat}
where
\begin{align*}
  \psi_l^m(\vec{k}, t)
  &= \braket{Y_l^m | \tilde{\psi}}
\end{align*}
is the coefficient  corresponding to spherical harmonics $Y_l^m $.

\begin{figure}
  \centering
 \includegraphics[width=0.95\linewidth]{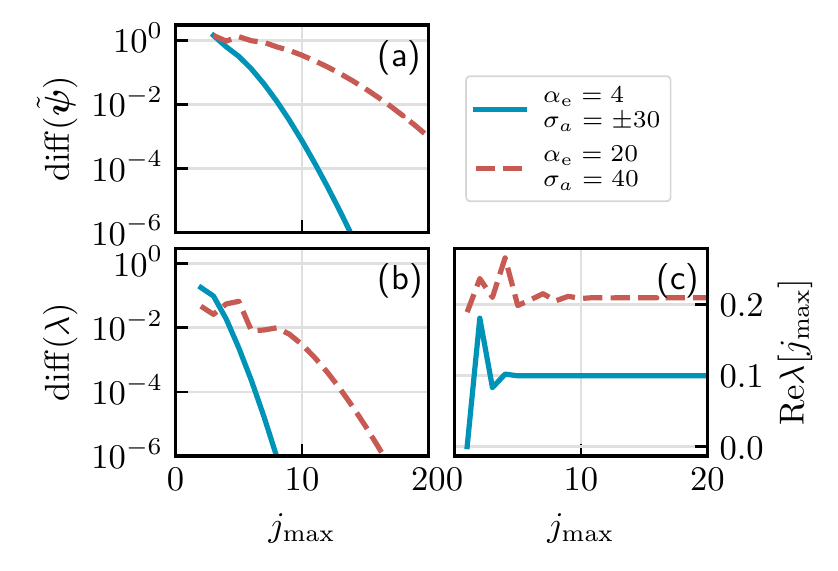}
  \caption{%
      (a) Magnitude of change of the corresponding eigenvectors (in the basis of spherical harmonics) with respect to a truncation order \(j_{\max}\),
    \(\mathrm{diff}(\vec{\tilde{\psi}}, j_{\max}) = \|{\vec{\tilde{\psi}}[j_{\max}] - \vec{\tilde{\psi}}[j_{\max}-1]}\|\). 
    The remaining parameters are fixed to \(\sigma_{\text{m}} = 0.4 \alpha_{\text{e}}, d_t = \num{3e-6}\).
    (b) Change of the largest eigenvalues \(\mathrm{diff}(\lambda, j_{\max}) = |\lambda[j_{\max}] - \lambda[j_{\max}-1]|\) as a function of the number of included modes \(j_{\max}\) on a logarithmic scale. 
    (c) The largest growth rate \(\Re \lambda_{\max}[j_{\max}]\) as a function of   truncation order \(j_{\max}\).
    }%
  \label{fig:hd_truncation_error}
\end{figure}

After substituting Eq.~\eqref{eq:harmonic_expand} into Eq.~\eqref{eq:hd_eigenvalue_problem} and applying the orthogonality condition Eq.~\eqref{eq:ortho}, the eigenvalue problem for the mode shape \(\tilde{\psi}(\vec{k}, \vec{n})\) reduces into an algebraic eigenvalue problem for the vector $\ket{\tilde{\psi}}$ whose components are given by the harmonic amplitudes \(\psi_l^h\):
\begin{equation}
 \sum_{j=0}^{\infty} \sum_{m=-j}^{j} L_{jl}^{mh} \psi_j^m = \lambda \psi_l^h
  \label{eq:algebraic_eigenvalue}
\end{equation}
in which \(L_{jl}^{mh} \equiv \braket{Y_l^h | \mathds{L}( Y_j^m)}\).
Expanding the operator \(\mathds{L}\) defined by Eq.~\eqref{eq:hd_perturbation_operator} on the spherical harmonics basis generates terms which are products of two spherical harmonics. The product can in general be written as the following linear combination of spherical harmonics
\begin{align}
  Y_{j_1}^{m_1}(\theta, \phi) Y_{j_2}^{m_2} (\theta, \phi)
  &= \sum_{j_3, m_3}
    \sqrt{\frac{(2 j_1 + 1) (2 j_2 + 1)}{4 \pi (2 j_3 + 1)}} \nonumber \\
  &\times \braket{ j_1, 0, j_2, 0 | j_3, 0} \nonumber \\
  &\times \braket{j_1, m_1, j_2, m_2 | j_3, m_3} Y_{j_3}^{m_3}.
\end{align}
in which \( \braket{j_1, m_1, j_2, m_2 | j_3, m_3} \) are known as the Clebsch-Gordan coefficients and their values are tabulated~\cite{racah_theory_1942,edmonds_angular_2016} and included in common software packages and computer algebra applications such as Mathematica.
The tensor \(\ten{L}\) in Eq.~\eqref{eq:algebraic_eigenvalue}
is of infinite size, hampering further analytical progress. We solve the algebraic eigenvalue problem by truncating the sum at sufficiently large \(j=j_{\max}\)  such that the convergence of the dominant eigenvalues and eigenvectors are ensured.
The number of angular modes that have to be included for convergence depends on \(\alpha_e\) partly because of the growing number of modes needed to accurately represent the steady state for large \(\alpha_e\). Truncating the coefficient tensor \(\ten{L}\) introduces an error in the calculation of the eigensystem. However, the error gets progressively smaller and has rapid convergence when adding further modes.
The value of the largest growth rate \(\Re \lambda_{\max}\) and 
as a function of \(j_{\max}\) is plotted in Fig.~\ref{fig:hd_truncation_error} for two different values of \(\alpha_e\).
For \(\alpha_e=4\), we find that \(j_{\max}=5\) ( \(55\) angular modes) is sufficient to obtain a good convergence whereas for
\(\alpha_e=10\), at least  \(j_{\max}=10\) ( 210  angular modes) is required for a reasonable convergence.

\subsection{Linear stability of homogeneous polar steady state}
\label{sec:hd_linstab_results}

As discussed earlier, the linear stability of the homogeneous steady state \(\psi_0\) in equation~\eqref{eq:hd_steady_state} depends on four dimensionless
parameters \(d_t\), \(\alpha_{\text{e}} \propto \mu B\), \(\sigma_{\text{a}}\) and \(\sigma_{\text{e}}\). Additionally, the eigenvalue problem defined by Eqs.~\eqref{eq:hd_eigenvalue_problem} and~\eqref{eq:hd_perturbation_operator} and thus the stability of the steady state depends on the direction of the wavevector \(\hat{\vec{k}}\) with respect to the field direction as the external field breaks the rotational symmetry. However, the system still holds an axial symmetry around the \(\vec{B}\) axis. Hence, the direction of wavevector can be characterized by a single angle between the magnetic field and the wavevector \(\Theta_B = \cos^{-1}(\hat{\vec{k}} \cdot \hat{\vec{B}})\). For a given solvent viscosity and density of active particles, the experimentally tuneable parameters are the strengths of  activity and magnetic field. Therefore, we construct a stability diagram as a function of
\(\alpha_{\text{e}} \propto \mu B\) and \(\sigma_{\text{a}}\). We set \(d_t = 3 \times 10^{-6}\); chosen to be comparable to the parameter ranges relevant for magnetotactic bacteria~\cite{waisbord_destabilization_2016}
and  vary the magnetic stress concomitantly with \(\alpha_{\text{e}}\) as  \(\sigma_{\text{m}} =  0.4 \, \alpha_{\text{e}}\).

\begin{figure}
  \includegraphics[width=0.95\linewidth]{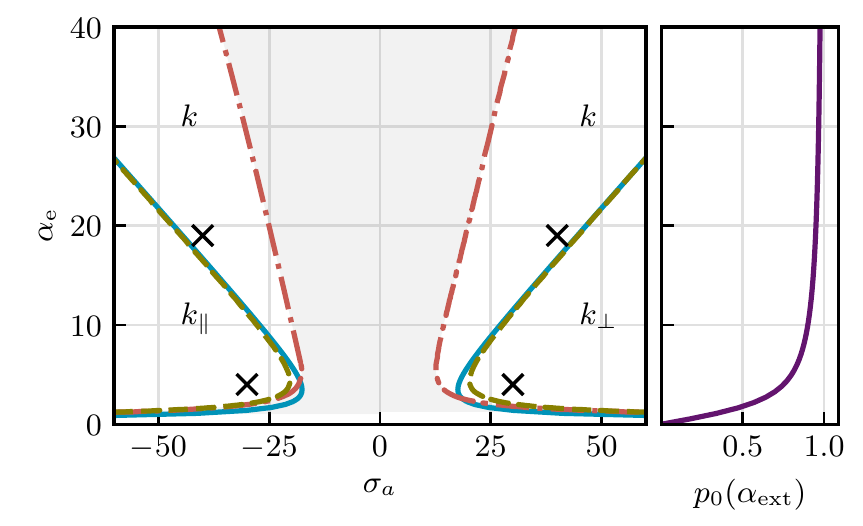}
  \caption{%
   Stability diagram of the steady state \(\psi_0(\alpha_{\text{e}})\) given by Eq.~\eqref{eq:hd_steady_state} as a function of the dimensionless active stress \( \sigma_{\text{a}} \) and alignment parameter \( \alpha_{\text{e}} \propto B\); while setting \(\sigma_{\text{m}} = 0.4\) \(\alpha_{\text{e}}\) and \( d_t = \num{3e-6}\).
    The borderline of neutral stability (red dash-dotted line) is calculated by finding \(\Re \lambda_{\max} (\vec{k})=0\). The dashed amber lines correspond to the cases where \(\Re \lambda_{\max} (\vec{k}_{||})=0\) for pushers and  \(\Re \lambda_{\max} (\vec{k}_{\bot})=0\) for pullers.
     The solid blue lines represent \(\Re \lambda_{\max} (\vec{k})=0\) based on the linear stability analysis of density and polarization fields  from truncated moment equations for wavevectors parallel and perpendicular to the magnetic field.
On the right side, the corresponding polarization \(p_0\) of the steady state \(\psi_0(\alpha_{\text{e}})\) is plotted.  For the points marked by crosses, the behavior of growth rate and  pattern formation are further discussed in the paper.
    }%
  \label{fig:hd_stab_dia}
\end{figure}

For a given set of parameters, \(\psi_0\) is unstable if the maximum growth rate is positive for at least one mode parametrized by \((k, \Theta_B)\). Based on the results of linear stability analysis, we divide the parameter space spanned by \((\sigma_{\text{a}}, \alpha )\) into stable and unstable regimes with respect to the steady state \(\psi_0\)~\cite{Koessel_2019}.
In the stable regime, the system evolves towards the steady state \(\psi_0\) and becomes stationary. In the unstable regime, even small fluctuations make the system depart from \(\psi_0\) towards a non-trivial dynamics. A line  of neutral stability, \emph{i.e.}, \(\Re \lambda_{\max}=0\) divides  the two regimes. In stability diagram of Fig.~\ref{fig:hd_stab_dia}, the red dashed-dotted lines represent the lines of neutral stability for pushers $\sigma_{\text{a}} <0$ and pullers $\sigma_{\text{a}} > 0$.  On the right panel, the mean polarization \(p_0(\alpha_{\text{e}})\) given by Eq.~\eqref{eq:hd_steadystate_polarization} is plotted, highlighting the dependency of the steady state \(\psi_0\) on \(\alpha_{\text{e}}\). The steady state \(\psi_0\) is stable for either of small activity \(|\sigma_{\text{a}}| \lesssim 20\) or a low external magnetic field \(\alpha_{\text{e}} \lesssim 0.5\).
In the case of small \(\sigma_{\text{a}}\), hydrodynamic interactions are too weak to destabilize the steady state. For a small \(\alpha_{\text{e}}\), the polarization \(p_0(\alpha_{\text{e}} \lesssim 0.5) \lesssim 0.3\) is rather weak and our system akin to an isotropic suspension of spherical swimmers remains stable.
For a sufficiently large active stress \(\sigma_{\text{a}} \gtrsim 20\) and a moderate external magnetic field strength, the homogeneous polar steady state becomes unstable. In this regime, the combined effect of sufficiently strong hydrodynamic interactions \(\propto \sigma_{\text{a}}\) and orientation fluctuations drive the system away from a uniformly aligned state. Interestingly, by further increasing the external magnetic field strength, the steady state becomes stable again, and we observe a reentrant hydrodynamic stability. Reentrant stability at strong external fields is a consequence of  magnetic torque overcoming the hydrodynamic torque. For given active stress amplitude $\sigma_{\text{a}}$, the active force $\vec{f}_{\text{a}}=\sigma_{\text{a}} \nabla \cdot \ten{Q}$ and its resulting hydrodynamic torque has an upper bound that can be overcome by the alignment torque  for sufficiently strong magnetic fields. As a result, the steady state becomes stable again.
\begin{figure}
  \centering
   \includegraphics[width=0.99\linewidth]{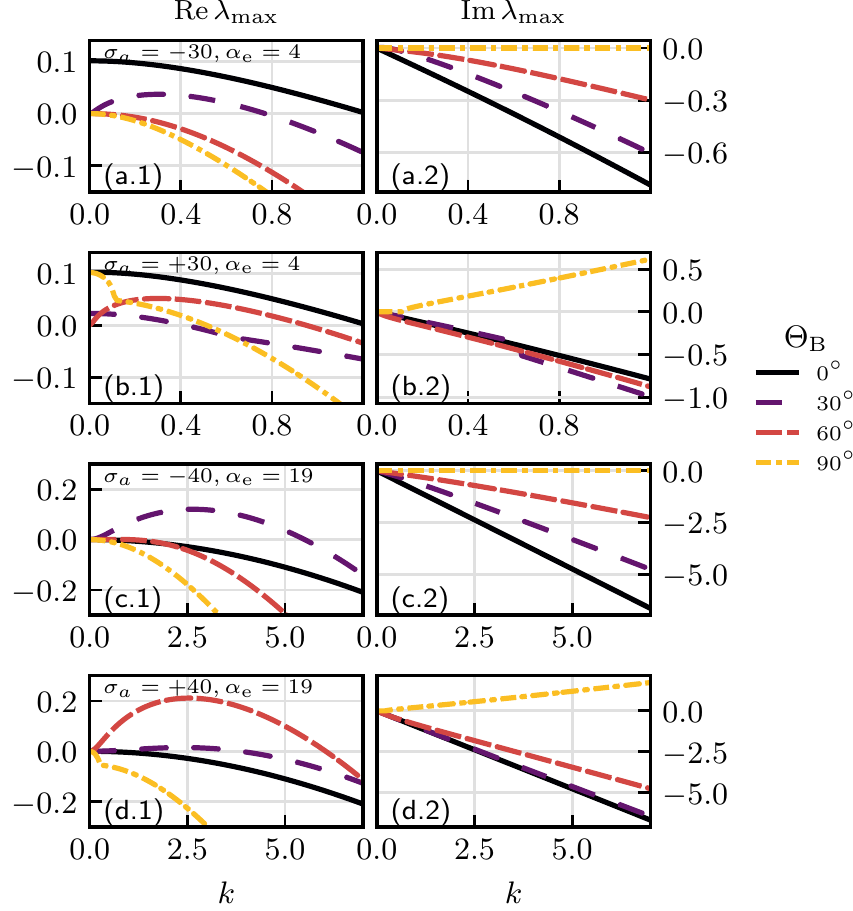}
  \caption{%
    Growth rate \(\Re \lambda_{\max}\) and oscillation frequency \(\Im \lambda_{\max}\) for different perturbation angle \(\Theta_B\). For
    (a) pusher at \(\sigma_{\text{a}} = -30, \alpha_{\text{e}} =4\),
    (b) puller at \(\sigma_{\text{a}} = 30, \alpha_{\text{e}} =4, \sigma_{\text{a}} = 30\), and
   (c) pusher at \(\sigma_{\text{a}} = -40, \alpha_{\text{e}} =19 \).
    (d) puller at \( \sigma_{\text{a}} = 40, \alpha_{\text{e}} =19 \).
  }%
  \label{fig:hd_growthrate_k}
\end{figure}

Analyzing the nature of instability in the unstable regions, we recognize four distinct types of instability.
To demonstrate their distinct nature, four representative points, corresponding to each type are picked out from the stability diagram. These points are marked by crosses in Fig.~\ref{fig:hd_stab_dia} and correspond to \((\sigma_{\text{a}}, \alpha_{\text{e}} )= (-30,4 ), (30, 4), (-40, 20)\) and \((40, 20)\) in the parameter space. Fig.~\ref{fig:hd_growthrate_k} shows the real and imaginary parts of the complex growth-rate with the largest real part, the so-called maximum growth rate (\(\Re \lambda_{\max} \)) and its oscillation frequency \(\Im \lambda_{\max}\), as a function of \(k = |\vec{k}|\) at various perturbation angles \(\Theta_B\) for each of the points. We note that the maximum growth rate strongly depends on the direction of the perturbation wavevector, \emph{i. e.}, \(\Theta_B\). For puller and pusher swimmers with equal strength of activity \(|\sigma_{\text{a}}|=30\) and magnetic field \(\alpha_{\text{e}} =4\),
long-wavelength perturbations \(k \to 0\) dominate the instabilities
and destabilize the homogeneous polar state \(\psi_0\). However, for pushers \(\sigma_{\text{a}}=-30\), fluctuations
in the direction of magnetic field grow fastest, whereas for pullers \(\sigma_{\text{a}}=30\) both perturbation directions parallel and perpendicular to \(\vec{B}\) predominate the system. Therefore, we expect distinct patterns of instabilities for pushers and pullers as confirmed by our non-linear simulations presented in Section~\ref{sec:pattern}.  Notably,  in both cases \(\Im \lambda_{\max}(k \to 0) \to 0\) which implies that the large wavelength fluctuations grow monotonically with time.
Interestingly, for pushers and pullers with stronger activity \(\sigma_{\text{a}}=\mp 40\) and much larger magnetic field \(\alpha_{\text{e}} =20\) but still in the unstable regime,  the wavenumber corresponding to the maximum growth-rate $k_{\text{max}}\approx 2.5$ is finite  and it occurs at  \(\Theta_B\approx 30^{\circ} \) for pushers and 
\(\Theta_B\approx 60^{\circ} \) for puller, featuring clearly different instability regimes.  In the case (d), \(\Im \lambda (\vec{k}_{\text{max}})\) of the wavevector with the maximum growth-rate is non-zero pointing to the oscillatory behavior of the predominant growth mode.
These  examples represent the four distinct types of instabilities observed: parallel and perpendicular orientational instabilities for pushers and pullers at moderate external field strengths, and more complex perturbation structures at higher field strength  featuring a finite characteristic wavenumber for the largest growth rate at an activity-dependent  angle intermediate between parallel and perpendicular directions.

\section{ Linear stability of  orientational moment equations}

\label{sec:hd_linear_stability_of_orientational_moments}
In this section, we take an alternative approach for investigating the linear stability of the homogeneous polar steady state. Instead of expanding the orientational part of the single particle distribution \(\psi \) in terms of spherical harmonics, as done in equation~\eqref{eq:th_psi_sh_expansion}, it equivalently
can be expanded in terms of dyadic products of the orientation vector \(\vec{n}\)~\cite{turzi_cartesian_2011, buckingham_angular_1967},
\begin{equation}
  \psi(\vec{x}, \vec{n}, t) = \sum_{j} \ten{M}(j, \vec{x}, \vec{n}) \odot^{j} \vec{n}^{\otimes j},
\end{equation}
where the \(l\)-fold dyadic product is denoted by
\begin{equation}
  \vec{n}^{\otimes l}
    = \underbrace{\vec{n} \otimes \ldots \otimes \vec{n}}_{l \, \mathrm{times}},
\end{equation}
and \(\odot^{l}\) denotes the \(l\)-fold contraction of two tensors,
\begin{equation*}
  {(\ten{A} \odot^{l} \ten{B})}^{i_1, \ldots, i_{n-l}} {}_{k_{l+1}, \ldots k_{m}},
  = A^{i_1, \ldots, i_{n-l}, j_{1}, \ldots j_{l}}
    B_{j_1, \ldots, j_{l}, k_{l+1}, \ldots k_{m}},
\end{equation*}
for tensors
\(
  \ten{A} \in T_0^{n}(\mathbb{R}^3)\) and \(\ten{B} \in T_{m}^{0}(\mathbb{R}^3)
\)
using Einstein summation convention. As shown in reference~\cite{turzi_cartesian_2011}, the coefficients  \( \ten{M}(j, \vec{x}, \vec{n}) \) are proportional to the orientational expectation values  of the symmetric and traceless (irreducible) part of
\(\vec{n}^{\otimes j}\),  where the orientational expectation value is defined as  \( {\left< \bullet \right>}_{\vec{n}} = \intn \psi \, \bullet \).
They are called the orientational moments of \(\psi \). 
Especially, the zeroth, first and second moments coincide with  the density, the polarization, and the nematic fields defined in equations~(\ref{eq:mom_density}-\ref{eq:mom_nematic}).
 Hence, the distribution \(\psi \) expanded in terms of the orientational moments reads
\begin{equation*}
  \psi(\vec{x}, \vec{n}, t) = \frac{1}{4 \pi} \left[
    \rho(\vec{x}, t)
    + 3 \vec{n} \cdot \vec{p}(\vec{x}, t)
    + \frac{15}{2} \vec{n} \vec{n} \cddot \ten{Q}(\vec{x}, t)
    + \cdots
  \right].
\end{equation*}
Truncating the moment expansion allows us  to manipulate the resulting terms algebraically and to find approximate analytical expressions for the linear stability analysis.  Evolution equations for each of the orientational moments can be  directly derived  by  taking moments of the Smoluchowski equation ~\eqref{eq:hd_smoluchowski}. The dynamical  moment equations for the first three moments are presented in the subsequent subsection.

\subsection{Equations of moments}

The time evolution of the density field \(\rho(\vec{x}, t)= {\left< 1 \right>}_{\vec{n}}\) can be derived by tracing out the angular dependency in equation~\eqref{eq:hd_smoluchowski}. Using the identity
\(
  \intn \slaplace \psi = 0
\),
its time evolution is given by
\begin{equation}
\mathcal{D}_t  \rho = - \nabla \cdot  \vec{p}  + d_t \laplace \rho.
  \label{eq:hd_density_time}
\end{equation}
where $\mathcal{D}_t \equiv   \partial_{t} +\vec{u}\cdot \nabla  $ represents the material derivative. This represents a convection-diffusion type of equation with a source term which originates from the divergence of polarization field.
Likewise, the time evolution of the polarization field \(\vec{p}(\vec{x},t)= {\left< \vec{n} \right>}_{\vec{n}}\) can be obtained by taking the first moment of equation~\eqref{eq:hd_smoluchowski}  and using the following identities.
\begin{equation}
  \intn \vec{n} \slaplace \psi
  = -2 \vec{p}
  \label{eq:hd_mom_integral_id1}
\end{equation}
and
\begin{equation}
  \intn
    \vec{n} \snabla \cdot ( \vec{v}_{\text{t}} \psi)
  = - \intn  \vec{v}_{\text{t}} \psi,
  \label{eq:hd_mom_integral_id2}
\end{equation}
in which \( \vec{v}_{\text{t}}\) represents any tangential vector field  on a sphere fulfilling \( \vec{v}_{\text{t}} \cdot \vec{n} = 0\), and in our context, it is given by the rotational drift velocity $\vec{v}_{\vec{n}}$ defined by equation~\eqref{eq:hd_smol_rot_flux}).
Consequently, the evolution  equation for the  polarization field is given by
\begin{align}
\mathcal{D}_t \vec{p} &=
   - \nabla \cdot \ten{Q} - \frac{1}{3} \nabla \rho \nonumber \\
 & -  \left( \ten{Q} - \frac{2}{3} \rho \mathds{1} \right) \cdot \alpha_{\text{e}} \hat{\vec{B}}+ \ten{W}\cdot \vec{p}
  - \ten{W}\cddot {\left< \vec{n}\vec{n}\vec{n} \right>}_{\vec{n}}  + d_t \laplace \vec{p}   - 2 \vec{p}.
\label{eq:hd_polarization_time}
\end{align}
which is again a convection-diffusion type of equation with a more complex source term including contributions from density gradient, polarization and the divergence of the nematic tensor field and terms arising from the interaction of the active particles with the local flow vorticity and  the magnetic field.

 Lastly, we obtain the time evolution of the nematic tensor field $\ten{Q}(\vec{x}, t)= {\left< \vec{n}\vec{n} - \mathds{1} / 3 \right>}_{\vec{n}}$   by integrating equation~\eqref{eq:hd_smoluchowski} with
\(
  \intn \vec{n}\vec{n} \, \bullet
\)
and using the following identities:
\begin{align}
  \snabla \vec{n}
  &= \vec{e}_\vartheta \vec{e}_\vartheta
     +\vec{e}_\varphi \vec{e}_\varphi
   = \mathds{1} - \vec{n}\vec{n}  \\
   \intn \vec{n}\vec{n} \snabla \cdot \vec{v_{\text{r}}} &= -\intn [\vec{n}\vec{v_{\text{r}}} + \vec{v_{\text{r}}}\vec{n}] \label{eq:hd_mom_integral_id3}\\
  \intn \vec{n}\vec{n} \slaplace \psi
  &= -2 \intn \vec{n} \psi \slaplace \vec{n}
    + 2 \intn \psi \snabla \vec{n} \nonumber \\
  &= -6 \ten{Q},
  \label{eq:hd_mom_integral_id4}
\end{align}
  The moment equation for $\ten{Q}$ is eventually given by
\begin{align}
  \mathcal{D}_t \ten{Q}
  &=    \frac{1}{3} \mathds{1} \nabla \cdot \vec{p}
  + \alpha_{\text{e}} (\vec{p} \hat{\vec{B}}  +\hat{\vec{B}}  \vec{p})
  -  \ten{W} \cdot \ten{Q}
  +  \ten{Q} \cdot \ten{W}
  - 6 \ten{Q} \nonumber \\
  &+ d_t \laplace \ten{Q}
  - \nabla \cdot {\left< \vec{n}\vec{n}\vec{n} \right>}_{\vec{n}}
  - 2 \alpha_{\text{e}} \hat{\vec{B}}  \cdot {\left< \vec{n}\vec{n}\vec{n} \right>}_{\vec{n}},
  \label{eq:hd_nematic_time}
\end{align}
in which we have used the  antisymmetric property of the  vorticity tensor \(\ten{W}\) to further simplify the equation. The dynamics of nematic tensor is directly affected by source terms steming from the divergence of polarization. Similar to nematodynamics equation of active nematics~\cite{active_nematics}, the dynamics of $\ten{Q}$ is strongly coupled to the flow velocity through the advection and vorticity terms. Moreover, additional terms appear due to coupling to the external field. 

As can be seen from the equations of moments Eq.~\eqref{eq:hd_density_time}, Eq.~\eqref{eq:hd_polarization_time}, and Eq.~\eqref{eq:hd_nematic_time}, they constitute a hierarchy of equations where each moment equation depends on higher moments. In order to proceed further,  we break this hierarchy  by introducing the following   closure relations which are compatible with a polar steady state
\begin{equation}
  \begin{split}
    \bar{\ten{Q}} &= \vec{p}\vec{p} - \frac{1}{3} \mathds{1} \\
    {\left< \vec{n}\vec{n}\vec{n} \right>}_{\vec{n}} &= 0,
  \end{split}.
  \label{eq:hd_momlinstab_closure}
\end{equation}
All higher order moments are neglected. These closure approximation is sometimes referred to the \emph{Hand}-closure~\cite{hand_theory_1962}.
 We will see in the next subsection that it generates perturbations that are structurally consistent with the results of  linear stability analysis of steady distribution function $\psi_0$ in Section~\ref{sec:hd_linear_stability_analysis}.

\subsection{Stability of moments}
After establishing the moment equations of the system, reduced to the density and polarization field, we proceed with their respective linear stability analysis employing the above closure relations. The homogeneous steady state solution of moment equations, Eqs.~\eqref{eq:hd_density_time} and ~\eqref{eq:hd_polarization_time} is given by $(\rho_0,\vec{p}_0)$ where
\begin{align}
   \rho_0 & = 1 \\
  \vec{p}_0 &=  \frac{\sqrt{4 \alpha_{\text{e}} ^2+9}-3}{2 \alpha_{\text{e}} }\hat{\vec{B}}.
\end{align}
which become  equivalent to ${\left< 1 \right>}_{\vec{n}}^{\psi_0}$ and ${\left< \vec{n} \right>}_{\vec{n}}^{\psi_0}=\frac{\alpha_{\text{e}} \coth \alpha_{\text{e}} - 1}{\alpha_{\text{e}} }$ (see Eq.~\eqref{eq:hd_steadystate_polarization}) for sufficiently small $\alpha_{\text{e}}$. 
Applying the closure relations Eq.~ \eqref{eq:hd_momlinstab_closure} yields $\ten{Q}_0=\vec{p}_0\vec{p}_0- \frac{1}{3} \mathds{1}$.

Next, we linearly disturb the steady state  by small perturbations of the form
\begin{align}
  \rho &=  \rho_0 + \varepsilon \rho_{p} \\
  \vec{p} &= \vec{p}_0 + \varepsilon \vec{p}_{p} \\
  \bar{\ten{Q}}
    &= \bar{\ten{Q}}_0  + \varepsilon \bar{\ten{Q}}_{p}
    = \left(\vec{p}_0 \vec{p}_0 - \frac{1}{3} \mathds{1} \right) + \varepsilon (\vec{p}_0 \vec{p}_{p} + \vec{p}_{p} \vec{p}_0) \\
  \vec{u} &= \vec{u}_0 + \varepsilon \vec{u}_{p},
\end{align}
in which \(0 < \varepsilon \ll 1\) and \(\vec{u}_0 = 0\), and substituting them into equations~\eqref{eq:hd_density_time} and ~\eqref{eq:hd_polarization_time}  the dynamics of the linearized perturbations in \(\mathcal{O}(\varepsilon)\) can be formulated as
\begin{alignat}{2}
  \partial_{t} \rho_{p} &&= &- \nabla \cdot \vec{p}_{p} + d_t \laplace \rho_{p} \label{eq:hd_mom_lin_rho} \\
  \partial_{t} \vec{p}_{p} &&=
    &\frac{2}{3} \rho_{p} \alpha_{\text{e}} \hat{\vec{B}}
    - \frac{1}{3} \nabla \rho_{p}
     + \ten{W}_{p} \cdot \vec{p}_0
    + d_t \laplace \vec{p}_{p}
    - 2 \vec{p}_{p} \nonumber \\
    &&&- \nabla \cdot \bar{\ten{Q}}_{p}
    - \bar{\ten{Q}}_{p} \cdot \alpha_{\text{e}} \hat{\vec{B}} .
    \label{eq:hd_mom_lin_pol}
\end{alignat}

Analogous to Section~\ref{sec:hd_linear_stability_analysis}, we make an eigenmode ansatz for the perturbations of the moments as below: 
\begin{align}
  \rho_{p}(\vec{k}; \vec{x}, t) &= \tilde{\rho}(\vec{k}) \, e^{\I \vec{k} \cdot \vec{x} + \lambda t} \\
  \vec{p}_{p}(\vec{k}; \vec{x}, t) &= \tilde{\vec{p}}(\vec{k}) \, e^{\I \vec{k} \cdot \vec{x} + \lambda t} \\
  \ten{Q}_{p}(\vec{k}; \vec{x}, t) &= \tilde{\ten{Q}}(\vec{k}) \, e^{\I \vec{k} \cdot \vec{x} + \lambda t}
               \equiv (\vec{p}_0 \tilde{\vec{p}}(\vec{k}) + \tilde{\vec{p}}(\vec{k}) \vec{p}_0) \, e^{\I \vec{k} \cdot \vec{x} + \lambda t} \\
  \vec{u}_{p}(\vec{k}; \vec{x}, t) &= \tilde{\vec{u}}(\vec{k}) \, e^{\I \vec{k} \cdot \vec{x} + \lambda t}.
\end{align}
Here, the flow field perturbation \(\tilde{\vec{u}}\) is mainly driven by \(\tilde{\ten{Q}}\) as
\begin{equation}
    \tilde{\vec{u}} = \frac{\I}{k^2}
    \left( \mathds{1} -\hat{\vec{k}} \hat{\vec{k}} \right)
    \cdot \sigma_{\text{a}} \tilde{\ten{Q}}
    \cdot \vec{k},
\end{equation}
in which \(\hat{\vec{k}} = k^{-1} \vec{k}\) and
\(
  \tilde{\ten{Q}} \equiv
  \vec{p}_0 \tilde{\vec{p}} + \tilde{\vec{p}} \vec{p}_0
\). Here, we have neglected the contribution of magnetic stress due to its small effect.
Substituting the eigenmode ansatz for perturbations back into the density and polarization equations, they transform into
\begin{alignat}{2}
  \lambda \tilde{\rho} &&=& - \I \vec{k} \cdot \tilde{\vec{p}} - d_t k^2 \tilde{\rho} \label{eq:hd_momlinstab_dens_perturbation}\\
  \lambda \tilde{\vec{p}} &&=
    &\frac{2}{3} \tilde{\rho} \, \alpha_{\text{e}} \hat{\vec{B}}
    - \frac{\I}{3} \vec{k} \tilde{\rho}
    + \tilde{\ten{W}} \cdot \vec{p}_0
    - d_t k^2 \, \tilde{\vec{p}}
    - 2 \tilde{\vec{p}} \nonumber \\
    &&&- \I \vec{k} \cdot \left( \vec{p}_0 \tilde{\vec{p}} + \tilde{\vec{p}} \vec{p}_0 \right)
    - \left( \vec{p}_0 \tilde{\vec{p}} + \tilde{\vec{p}} \vec{p}_0 \right) \cdot \alpha_{\text{e}} \hat{\vec{B}} ,
    \label{eq:hd_momlinstab_pol_perturbation}
\end{alignat}
 where again $\tilde{\ten{W}} = \frac{\I}{2} \left( \tilde{\vec{u}} \vec{k} - \vec{k} \tilde{\vec{u}} \right)$.
 
To analyze these equations, we consider  perturbations of polarization which are perpendicular to   the external field $\vec{B}=B  \hat{\vec{z}}$ and polarization \(\vec{p}_0\equiv p_0 \hat{\vec{z}} \). Without loss of generality, we assume the perturbation to be in the \(x\)-direction \emph{i.e.}, \(\tilde{\vec{p}} \equiv \tilde{p}(\vec{k}) \hat{\vec{x}}\). This assumption reduces the problem to 2D and moreover,
 \(\tilde{\vec{p}} \cdot \vec{p}_0 = 0\). It also implies that small linear perturbations perpendicular to  \(\vec{p}_0 \) practically  influence the orientation but not the magnitude of the polarization and allows us to investigate hydrodynamically induced orientational instabilities. Moreover, it naturally generates a nematic perturbation of the form \(\tilde{\vec{p}} \vec{p}_0 + \vec{p}_0 \tilde{\vec{p}}\)  which
 is compatible with the closure approximation given by Eq.~\eqref{eq:hd_momlinstab_closure}.  Physically, a polarization perturbation of this form corresponds to bend (for pushers) and splay (for pullers) deformations of the polarization field~\cite{ramaswamy_active-filament_2007}, to be discussed in the following Section.
 Setting \(\tilde{\vec{p}} \cdot \vec{p}_0 = 0\) and thus focusing on \(\tilde{\rho}\) and \(\tilde{p}_{\perp}\), the equations~\eqref{eq:hd_momlinstab_dens_perturbation} and~\eqref{eq:hd_momlinstab_pol_perturbation} can be written in the reduced form of
\begin{equation}
  \lambda \left(
  \begin{array}{c}
    \tilde{\rho} \\
    \tilde{p}_\perp
  \end{array}
  \right)
  =
  \ten{T}
  \cdot
  \left(%
    \begin{array}{c}
      \tilde{\rho} \\
      \tilde{p}_\perp
    \end{array}
  \right),
\end{equation}
with the operator \\
\resizebox{0.5\textwidth}{!}{$\ten{T} \equiv
  \begingroup 
  \setlength\arraycolsep{10pt}
  \begin{pmatrix}
    -d_t k^2 & -i k \sin \Theta_B   \\
    -\frac{1}{3} \I k \sin \Theta_B & -2 - d_t k^2 - \I p_0 \, k \, \cos \Theta_B -\alpha_{\text{e}} p_0 -\frac{1}{2} p_0^2 \sigma_{\text{a}} \cos 2 \Theta_B
  \end{pmatrix}
  \endgroup.$}
Here, \(
  \vec{k} = k \sin(\Theta_B) \vec{\hat{e}}_x
          + k \cos(\Theta_B) \vec{\hat{e}}_z
\)
was used, where \( \Theta_B \) denotes the angle of the direction of the perturbation with respect to the external field axis pointing in the $z$-direction, \(\Theta_B = \angle(\vec{k}, \vec{B} )\). The eigenvalues can be found analytically\footnote{%
If in doubt regarding dropping the parallel component of \(\tilde{\vec{p}}\) in the calculation, it should be noted the same result is obtained at \(k=0\) when including it.
}.

Solving this eigenvalue problem, we find that the largest eigenvalue can be found at \(k \to 0\), in agreement with the findings of Section~\ref{sec:hd_linstab_results} for the cases (a) and (b) in Fig.~\ref{fig:hd_growthrate_k}. Therefore, for stability regimes (a) and (b), it is sufficient to restrict the analysis to parallel and perpendicular modes of truncated moment equations to assess the stability of a uniform polarization field. The \(k \to 0\) eigenvalues read
\begin{equation}
  \lambda_{1,2} =
  \frac{1}{4}
  (a \pm |a|)
  \label{eq:hd_evals_momentmatrix}
\end{equation}
with
\(
  a =
    -4
    -2 \alpha_{\text{e}} p_0
    -p_0^2 \sigma_{\text{a}} \cos 2 \Theta_B
\).
The largest eigenvalue is given by \(\lambda_{\text{max}} = \max(0, a/2)\) (and the smallest by \(\min(0, a/2)\)). However, simulations show that the perturbation associated with the \(\lambda=0\) can be considered stable. On the other hand, the eigenvalue corresponding to the mode that becomes unstable, \emph{i.e.}, the one having a change of sign in its growth rate, can be constructed by combining the non-zero parts of eigenvalues into one, yielding
\begin{equation}
    \lambda = \frac{a}{2} =
    -2
    -\alpha_{\text{e}} p_0
    -\frac{1}{2} p_0^2 \sigma_{\text{a}} \cos 2 \Theta_B.
    \label{eq:hd_momlinstab_lambdamax}
\end{equation}

\begin{figure}
    \centering
     \includegraphics[width=0.95\linewidth]{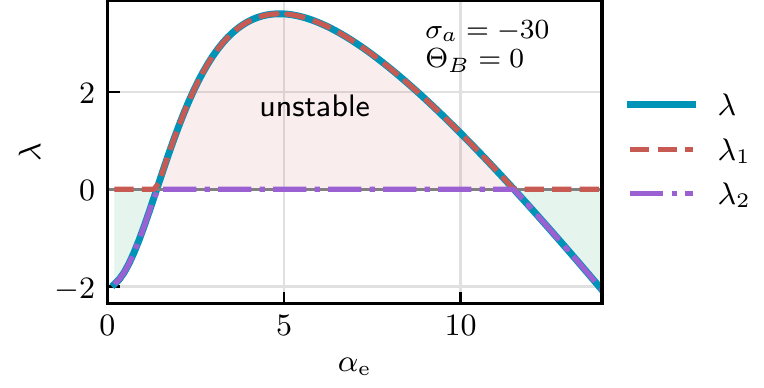}
    \caption{%
        The growth rates \(\Re \lambda\) against the alignment parameter \(\alpha_{\text{e}} \) of a linear perturbation of the polarization field, equation~\eqref{eq:hd_momlinstab_lambdamax}. This illustrates the re-entrant stability upon increase of magnetic field. The blue line is the eigenvalue of a mode crossing over from negative to positive and back to negative growth rate. It is constructed from the eigenvalues \(\lambda_{1,2}\) defined in equation~\eqref{eq:hd_evals_momentmatrix}. 
    }%
    \label{fig:hd_momlinstab_lambdamax}
\end{figure}

The dependence of the non-zero eigenvalue \(\lambda \) on the  alignment parameter \(\alpha_{\text{e}} \) is plotted in Fig.~\ref{fig:hd_momlinstab_lambdamax}.
This result demonstrates that the simplified approach is sufficient to recover the re-entrant stability obtained  earlier based on the full linear stability analysis of the steady state.

The line of neutral stability can be found by solving \(\lambda = 0\) for \(\sigma_{\text{a}}\):
\begin{equation}
    \sigma_{\text{a}}^0(\alpha_{\text{e}} ) =  -\frac{2 p_0(\alpha_{\text{e}} ) \, \alpha_{\text{e}} + 4}{{p_0(\alpha_{\text{e}} )}^2  \, \cos 2 \Theta_B }.
    \label{eq:hd_momlinstab_null}
\end{equation}
It is shown in Fig.~\ref{fig:hd_stab_dia} in direct comparison to the result of linear stability analysis of the steady state distribution function $\psi_0$.
The line of neutral stability based on the stability of the first two moments (blue solid line) nearly coincides with the results of the parallel and perpendicular perturbations obtained from the linear stability analysis of \(\psi_0\) discussed in Section~\ref{sec:hd_linstab_results} (red dashed line). For larger external field strengths, the full analysis reveals additional unstable modes that do not fall into the same scheme. For the regions between the red dash-dotted lines and dashed amber lines in Fig.~\ref{fig:hd_stab_dia}), a wider spectrum of orientational modes contribute to the instability that are not captured by the closure approximation of equation~\eqref{eq:hd_momlinstab_closure}. Additional moments would be required to obtain a complete description.

\section{NONLINEAR DYNAMICS SIMULATIONS}%
\label{sec:pattern}
The \emph{linear} stability analysis predicts the stability of a given steady state and provides us with a qualitative insight into the dynamics near it. However, as the system departs from the initial steady state, non-linearities prevail the dynamics and the linearized equations  fail to describe the dynamics correctly. Therefore, it is necessary to investigate  solution of the full non-linear equation~\eqref{eq:hd_smoluchowski}.  Below, we first outline our methodology for solving the full non-linear Smoluchowski equations coupled to the Stokes flow. Then, we discuss the  pattern formation  emerging from  the long-time dynamics of active magnetic swimmers in the external field.

\subsection{Numerical simulation method}
 Our methodology consists of  a hybrid stochastic particle based sampling method for obtaining \( \Psi(\vec{x},\vec{n},t) \) in the Smoluchowski equation  with periodic boundary conditions. It is based on integrating  coupled translational and rotational Langevin equations, which are the counterpart of the Smoluchowski equation. They are given by: 
\begin{align}
  \dot{\vec{x}} &= \vec{v}_{\vec{x}}(\psi (\vec{x}, \vec{n}, t), \vec{x}, \vec{n}, t) + \sqrt{2 d_{\text{t}}} \ten{\Gamma}(t)
  \label{eq:met_general_langevin_psi1_translation} \\
  \dot{\vec{n}} &= \vec{v}_{\vec{n}}(\psi (\vec{x}, \vec{n}, t), \vec{x}, \vec{n}, t) + \sqrt{2} \ten{\Lambda}(t) \times  \vec{n},
  \label{eq:met_general_langevin_psi1_rotation}
\end{align}
where $\vec{v}_{\vec{x}}$ and $\vec{v}_{\vec{n}}$ are defined by equations \eqref{eq:v_t} and \eqref{eq:v_r}.  \(\ten{\Gamma}\) and \(\ten{\Lambda}\) represent the stochastic force and torques with the following statistical properties:
\begin{eqnarray}
 \langle \ten{\Gamma}(t) \rangle =0&, \, \langle \Gamma_i(t)  \Gamma _j(t') \rangle =\delta_{ij} \delta(t-t'), \\
 \langle \ten{\Lambda}(t) \rangle =0&, \,  \langle  \Lambda_i(t)  \Lambda_j(t') \rangle =\delta_{ij} \delta(t-t')
 \end{eqnarray}

\begin{figure}
  \centering
  \includegraphics[width=0.49\textwidth]{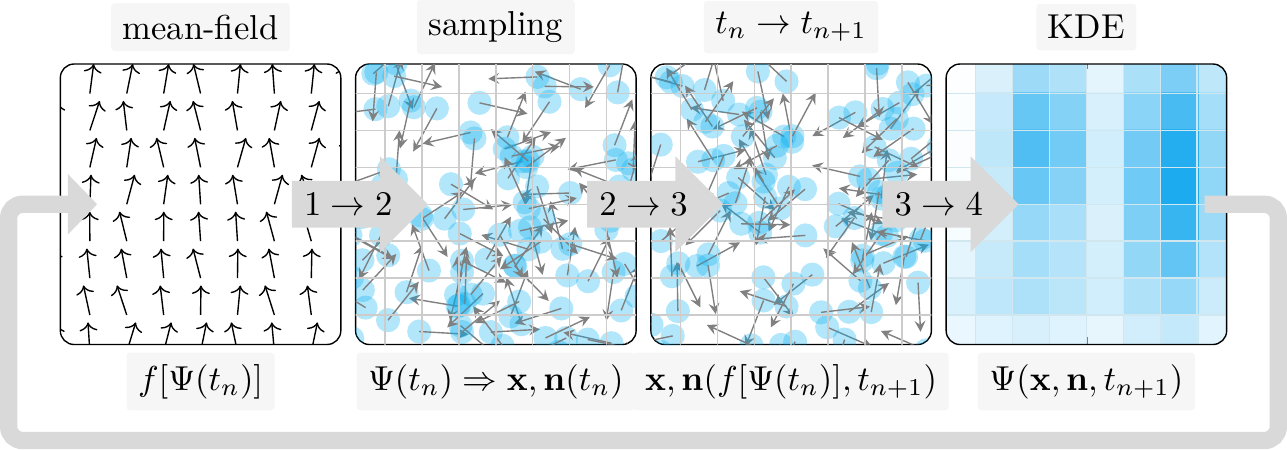}
  \caption{%
    Flow diagram illustrating the algorithm of \emph{Stochastic Sampling} method used to solve full non-linear Smoluchowski equation coupled to the mean-field Stokes flow. Given a probability density function (PDF) \(\Psi\), the mean flow field can be calculated by evaluating the stress profile based on  Eq.~\eqref{eq:mean_stress}. Then, the PDF can be sampled to obtain particles which are integrated using the interaction field. Employing the updated sample configuration the PDF can be estimated by a Kernel Density Estimation (KDE). By iterating the process, the Smoluchowski equation describing the dynamics of the PDF can be integrated in time.
  }%
  \label{fig:th_stochastic_sampling}
\end{figure}

 Within our theory,  direct inter-particle dependencies are replaced by mean-field interactions. Consequently, once the mean-field  stress profile and the resulting flow is computed from the distribution function, different initial value problems for a given particle can be simulated independently of each other. This realization is the basis of our numerical method that we dub  ``Stochastic Sampling'' method.  Fig. \ref{fig:th_stochastic_sampling} summarizes the flow diagram of our method. A detailed  description of the methodology can be found in reference ~\cite{Koessel_thesis}.

To solve these equations numerically,  we employ  the Euler forward integration scheme based on the Itô interpretation of noise.
  For every time step, we integrate the corresponding Langevin stochastic differential equations for the positions and orientations of a large number (\(10^6 - 10^8\)) of independent and randomly initialized test particles.
 These sample configurations provide us with sufficient statistics to  infer a realization of the distribution function $\psi(\vec{x},\vec{n},t)$ by using a kernel density estimation method. From the estimated distribution function, we compute the stress profile in the fluid. We use a spectral method based on the decomposition of $\vec{u}$ into the Fourier modes for solving the Stokes equation. In the Fourier domain, the flow field  is obtained as   
   \( \vec{u} _p^{\mathcal{F}}(\vec{k})=  \frac{1}{k^2} (\mathds{1} - \hat{\vec{k}} \hat{\vec{k}})\cdot  (\I \vec{k} \cdot \ten{\sigma}_p^{\mathcal{F}}) \)
   in which \( \ten{\sigma}_p^{\mathcal{F}} \) is the Fourier transform of the stress tensor.  Given the stress profile,  the  flow field  in terms of its Fourier modes on a periodic lattice is obtained. Then, it is Fourier transformed back to the real space.  Eventually, \(\vec{u}(\vec{x},t)\) is fed back into the next integration time step for the Langevin equations.
   
   In the reported numerical simulations, we use a grid of 100 lattice points with box dimensions of \(5\, x_c\) for each of the spatial coordinates, and 24 and 16 points for the spherical polar and azimuthal orientational coordinates \(\theta\) and \(\phi\) in \(\vec{n}(\theta, \phi)\). This
choice of box dimension ensures that the initial perturbation
spans both unstable and stable modes for all the four instability regimes presented in Fig.~\ref{fig:hd_growthrate_k}.
The simulations, conducted in a  box size of \(V_{\text{sim}}={(5\, x_c)}^3\), are initialized with the homogeneous polar steady state \(\psi_0\), given by equation~\eqref{eq:hd_steady_state} for different system parameters. As in the previous sections, the translational diffusion is fixed to \(d_t = \num{3e-6} \), the alignment stress magnitude is set to \(\sigma_{\text{m}} = 0.4 \alpha_{\text{e}}\), while the external magnetic field and the active stress magnitude are varied through the alignment parameter \(\alpha_{\text{e}}\) and the active stress amplitude \(\sigma_{\text{a}}\).

\begin{figure}[ht]
  \centering
   \includegraphics[width=0.99\linewidth]{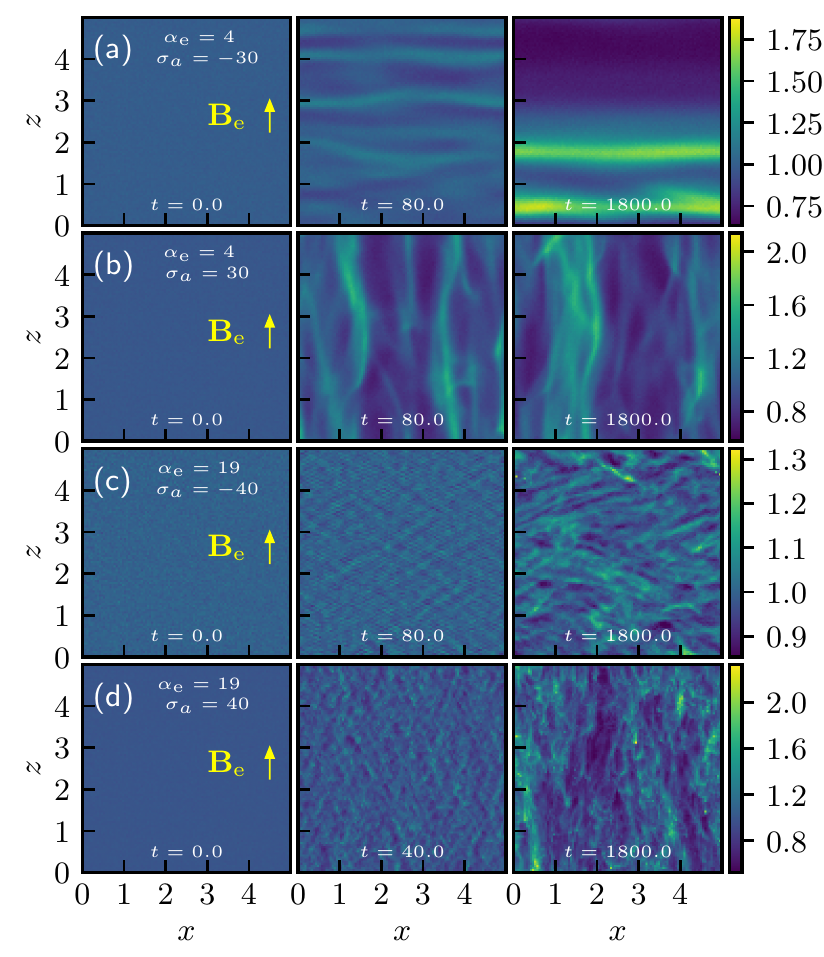}
     \caption{%
   Representative  snapshots of density field projections averaged along the y-axis from 3D non-linear simulations at different time steps, as shown on the snapshots, for magnetic swimmers with activity strength and alignment parameter values  (a)  \(\sigma_{\text{a}}=-30\) and  \(\alpha_{\text{e}}=4\)
   (b) \(\sigma_{\text{a}}=30\) and  \(\alpha_{\text{e}}=4\), (c) \(\sigma_{\text{a}}=-40\) and  \(\alpha_{\text{e}}=19\) and (d) \(\sigma_{\text{a}}=40\) and  \(\alpha_{\text{e}}=19\). 
    The colors encode the probability density integrated in the y-direction \(\bar{\rho}(x, z) = \Delta y{\sum}_y \, \rho(x, y, z)\). In all simulations, the translational diffusion is fixed to \(d_t = \num{3e-6}\) and the alignment stress is varied along the external field as \(\sigma_{\text{m}} = 0.4 \alpha_{\text{e}}\).  }%
  \label{fig:hd_num_density}
\end{figure}
\begin{figure*}
  \centering
 \includegraphics[width=0.95\linewidth]{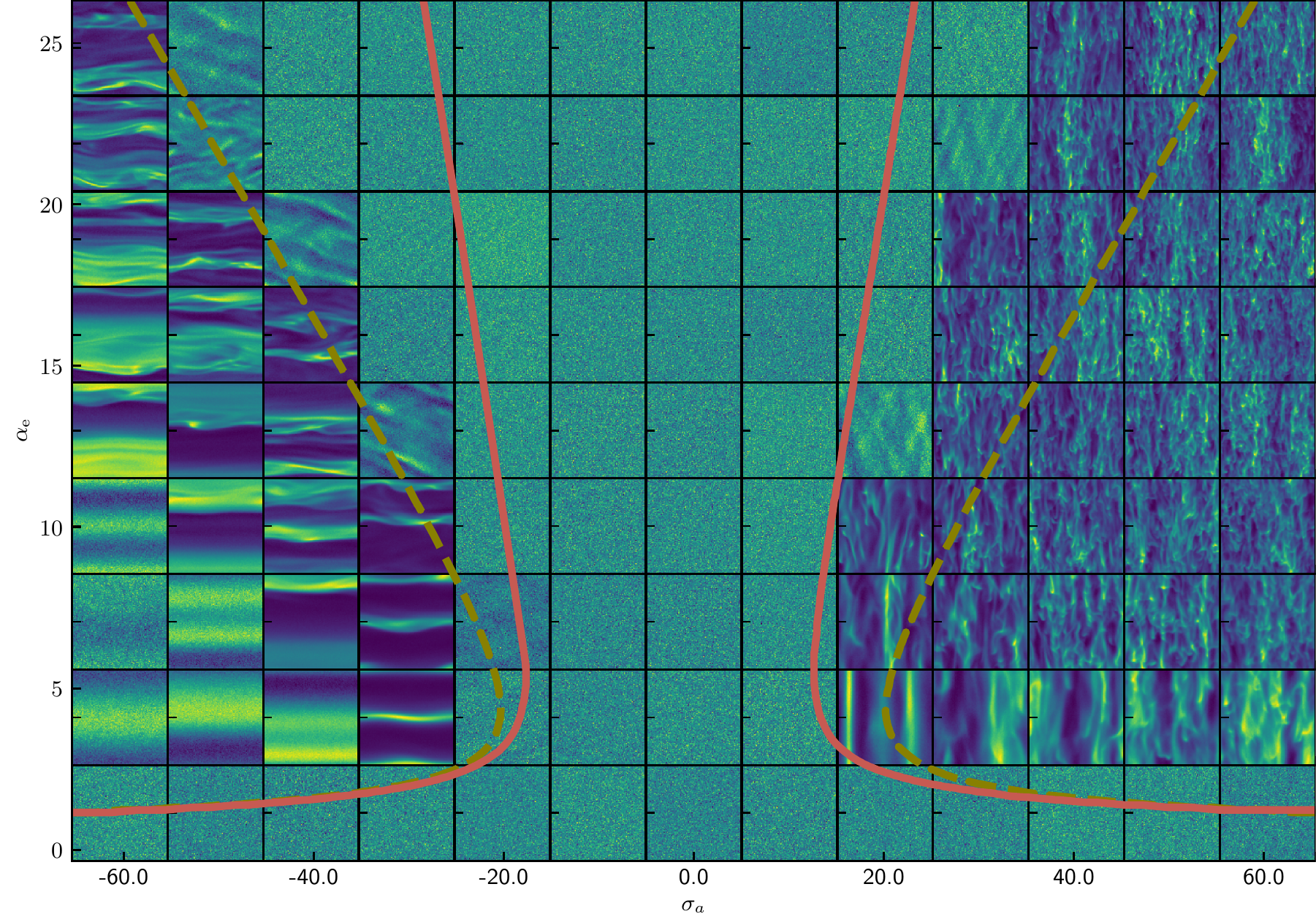}
  \caption{%
    Density projections for different \((\alpha_{\text{e}},\sigma_{\text{a}})\)  values   superposed by the lines of neutral stability from linear stability analysis. The red lines marks the onset of  instability calculated by the linear stability analysis probing all possible orientations of wavevector. The amber dashed lines correspond to predictions for neutral stability (\(\Re \lambda_{\max} (\vec{k})=0 \)) when considering only perturbations parallel $\vec{k}_{||}$ (pushers) and perpendicular $\vec{k}_{\bot}$ (pullers) to the external field. The translational diffusion coefficient is fixed to \(d_t = \num{3e-6}\) and the alignment stress is  varied with the external field as \(\sigma_{\text{m}} = 0.4 \, \alpha_{\text{e}}\).
  }%
  \label{fig:hd_num_density_overview}
\end{figure*}

\subsection{Pattern formation and nature of instabilities}
Starting from a spatially homogeneous polar state $\psi_0(\alpha_{\text{e}})$ given by Eq. \eqref{eq:hd_steady_state}, we evolve $\psi$ and $\vec{u}$ for each state point characterized by the \((\alpha_{\text{e}},\sigma_{\text{a}})\) pair by employing the method outlined above. 
For the unstable state points, $\psi$ departs from $\psi_0$ significantly whereas for the stable points the system converges towards $\psi_0$ even starting from an initial uniform isotropic state. The snapshots of Fig.~\ref{fig:hd_num_density}  depict  the time evolution of density field, projected to 2D by averaging  along the y-axis, for   \((\alpha_{\text{e}},\sigma_{\text{a}})\) values corresponding to 
the points marked by crosses in Fig.~\ref{fig:hd_stab_dia}.  We observe a general trend that a uniform density profile becomes unstable towards density fluctuations.
Over time, small-scale fluctuations disappear and the  the field profiles become smoother owing to diffusion. Only predominant fluctuations at wavelengths of the order of the box size persist. As a consequence, smooth non-uniform density, polarization and flow fields develop. At long times, the configuration of the active suspensions is not steady but constantly fluctuates in time. The distribution of swimmer  orientation  appears to converge towards  a dynamical steady state which depends on \( \alpha_{\text{e}}\) and \(\sigma_{\text{a}}\), leading to a constant average polarization in time ~\cite{Koessel_2019}. In contrast, the density fields exhibit distinct spatial patterns  for different instability regimes which  keep evolving and reorganizing.

Using the results from the non-linear dynamics simulations,  we assess the validity of  the phase diagram of Fig.~\ref{fig:hd_stab_dia} predicted by the linear stability of $\psi_0(\alpha_{\text{e}})$.  Fig.~\ref{fig:hd_num_density_overview} presents an overview of   density field  projections into the \(x\)-\(z\)-plane, at a late time t=1800, after the instability has already established itself, for different values of the active stress \(\sigma_{\text{a}}\) and the alignment parameter \(\alpha_{\text{e}}\).  To compare with the linear stability analysis predictions, we have plotted the lines of neutral stability (red lines), for which the largest growth rate is zero, \emph{i.e.}, \(\Re \lambda_{\max} = 0\).  Additionally, the dashed amber lines depict the  borderlines beyond which  instability is governed by parallel  and perpendicular perturbations for negative (pusher) and positive (puller) \(\sigma_{\text{a}}\), respectively. Comparing the predictions of the linear stability analysis with results of simulations for different activity and magnetic field strengths, we find excellent agreement.   \(\psi_0(\alpha_{\text{e}})\) is stable for the (\(\sigma_{\text{a}} \), \(\alpha_{\text{e}}\))  values where  density profile remains homogenous, whereas \(\psi\) evolves towards an inhomogeneous time-dependent density profile for the regions that are unstable according to the  linear stability analysis.

For moderate external magnetic field strengths  \(\propto \alpha_{\text{e}}\) and moderate activities \(\propto \sigma_{\text{a}}\), corresponding to the unstable regions beyond the amber lines, patterns with the most distinct characteristics appear. The higher the external field, the patterns display finer structures (of higher spatial frequency) in the density profile suggesting a the predominance of the characteristic length scale associated with the external field varying inversely with magnetic field strength. Indeed, based on dimensional analysis, one can identify a length scale $\ell_e\propto \sqrt{D_t \eta/ \varrho_m \mu B}\equiv \sqrt{D_t \eta /\Sigma_{\text{m}}}$.
In these regions, the basic characteristics seem to be conserved, with a band-like structure for pushers and a pillar-like structure for pullers. In the intermediate regions between the red and amber lines, the perturbative mode structure of the instabilities differs from the region beyond the amber line, as discussed in section \ref{sec:hd_linstab_results} and particularly in Fig.~\ref{fig:hd_stab_dia}. For pushers, we observe band-like density structures which are not perpendicular to the magnetic field. Pullers in the intermediate region mainly produce a similar density profile as in the region beyond amber line  but with  finer density structures. Notably, at higher magnetic fields, \emph{i.e.}, larger $\alpha_{\text{e}}$, we observe  hatch-like patterns of density  with finite width, which are not parallel to the magnetic field.  In general, as expected, the instabilities near the line of neutral stability are rather weak and the resulting dynamics only depart very slowly from the steady state.
  Consistent with the linear stability analysis prediction, we observe distinct patterns for pushers and pullers.    More insight into instabilities and their underlying mechanism can be gained by investigating the  polarization and self-generated flow fields.  In the following, we discuss the prominent features of the long-time density, polarization and flow fields of each of the representative points discussed in Fig.~\ref{fig:hd_growthrate_k} and Fig.~\ref{fig:hd_num_density} corresponding to four  distinct instability regimes.
\begin{figure}
  \centering
   \includegraphics[width=0.95\columnwidth]{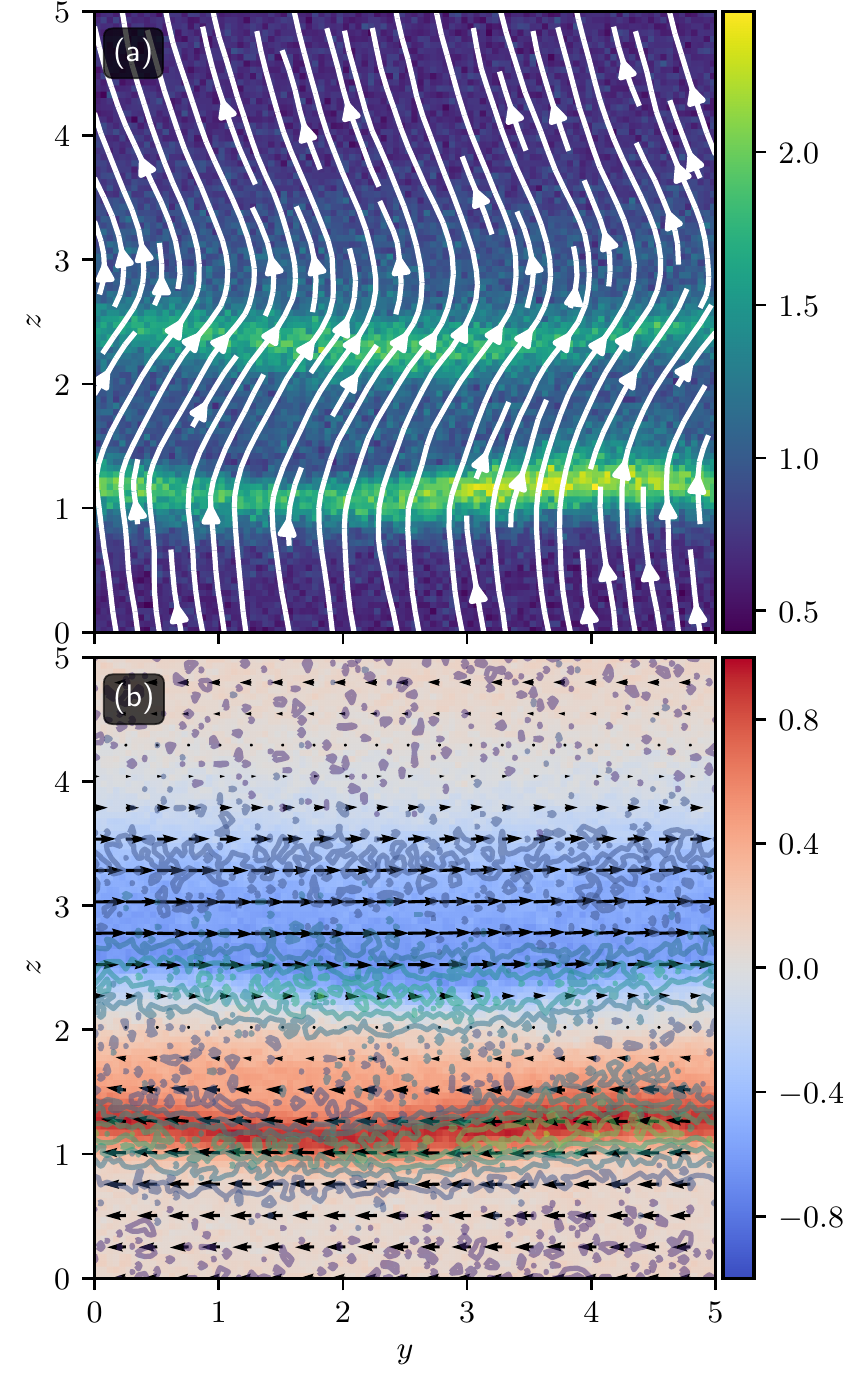}
  \caption{%
    (a)  Streamlines of the polarization field $(p_y,p_z)$ of a slice in the \(y\)-\(z\)-plane  at $x=0.75$,   demonstrating a characteristic bend fluctuations for a suspension of pushers with dimensionless active stress \(\sigma_{\text{a}} = -30\) and alignment parameter \(\alpha_{\text{e}} = 4\) in a magnetic field pointing in \(z\)-direction. The color encoded local density  \(\rho / V \times 10^6 \) is shown in the background. 
   (b) The corresponding flow field $(u_y,u_z)$ is represented as vector arrows where the length of the vector is weighted by its magnitude. The flow vorticity, responsible for the hydrodynamically induced particle rotation, is color encoded in the background. Red colors correspond to a vorticity vector pointing out of the plane (counter clockwise rotation).  Lines of constant  density are overlaid as  contour lines on top to guide the eyes. 
      }%
  \label{fig:hd_push_yz_slice}
\end{figure}

\subsubsection{Instability regime (a): traveling bands} 
 
The snapshots in  Fig.~\ref{fig:hd_num_density} (a)  present pattern formation for pushers with dimensionless active stress \(\sigma_{\text{a}} =-30\) and alignment parameter \(\alpha_{\text{e}}=4\). They correspond to the  panels (a.1) and (a.2) of Fig.~\ref{fig:hd_growthrate_k}, where the linear stability analysis predicts the prevalence of the long wavelength instabilities of wavevectors parallel to the magnetic field. At late stages of the simulations,  we observe density modulations in the direction parallel to $\vec{B}$, confirming the development of such long wavelength instabilities. The swimmers concentrate in bands perpendicular to the magnetic field  spanning the whole transverse dimension of the box while traveling collectively in the field direction. 
  To better understand  the origin of these hydrodynamic instabilities, we investigate the corresponding polarization and flow fields. We first focus on a 2D slice of the sample.  Fig.~\ref{fig:hd_push_yz_slice}(a) displays the polarization  field  superimposed by the density field at a late stage $t=1800$.
  In  Fig.~\ref{fig:hd_push_yz_slice} (b), the corresponding flow and vorticity fields are presented.  We note that concomitant with density modulations, the polarization and flow fields also become non-uniform.  The polarization streamlines begin to deflect from straight lines forming bend-like deformations.  Such distortions can be understood in terms of bend instability of polarization field similar to  those observed in liquid crystals.

  Bend fluctuationsconsist of small polarization perturbations which are perpendicular to \(\vec{p}_0\) while their magnitude is modulated in the direction parallel to \(\vec{p}_0 \parallel \hat{\vec{B}}\equiv \hat{\vec{z}}\), {\it i.~e.}, $\vec{p}_p=\tilde{\vec{p}}_{\bot}  \exp( \I  k_{z} z )$~\cite{LC_Prost,ramaswamy_active-filament_2007}.  Such distortions  increase the density  in volumes of negative divergence \(\nabla \cdot \vec{p} < 0\), as expected based on time evolution of density given by Eq.~\eqref{eq:hd_density_time}. As a result, pushers form dense layers  perpendicular to \(\hat{\vec{B}}\) that migrate parallel to the magnetic field.  
    Bend-like distortions also generate a  position-dependent active stress $\propto \ten{Q}$ that results  in a net active force density  in the fluid  given by  $\vec{f}_{\text{a}} \approx \sigma_{\text{a}} \nabla \cdot \left( \vec{p} \vec{p} - \frac{1}{3} \mathds{1} \right)$.   The active force density leads to alternating flow layers perpendicular to the magnetic field as can be observed from Fig.~\ref{fig:hd_push_yz_slice}(b). The ensuing vorticity field, encoded by background color, is also modulated in  a similar fashion. According to the Faxen's second law \cite{dhont_introduction_1996},  the spherical microswimmers is affected by the hydrodynamically induced torque \(\vec{M}_{\text{HD}} \propto \frac{1}{2} \nabla \times \vec{u} \) due to the flow vorticity,  which rotates the swimmers further away from the magnetic field axis. Thus, the self-generated flow amplifies the bend distortions  and renders  a uniform homogenous polar phase unstable. This self-amplification would lead to a highly unstable feedback loop if the external torque \(\vec{M}_{\mathrm{B}} \propto \alpha_{\text{e}} \vec{p} \times \hat{\vec{B}} \) would not eventually counterbalance the hydrodynamic torque. The competition between the alignment and hydrodynamic torques continues until  they almost balance each other, hindering further growth of instabilities. As a result, fairly stable patterns at dynamic equilibrium are established. 
\begin{figure}
  \centering
  \includegraphics[width=0.45\columnwidth]{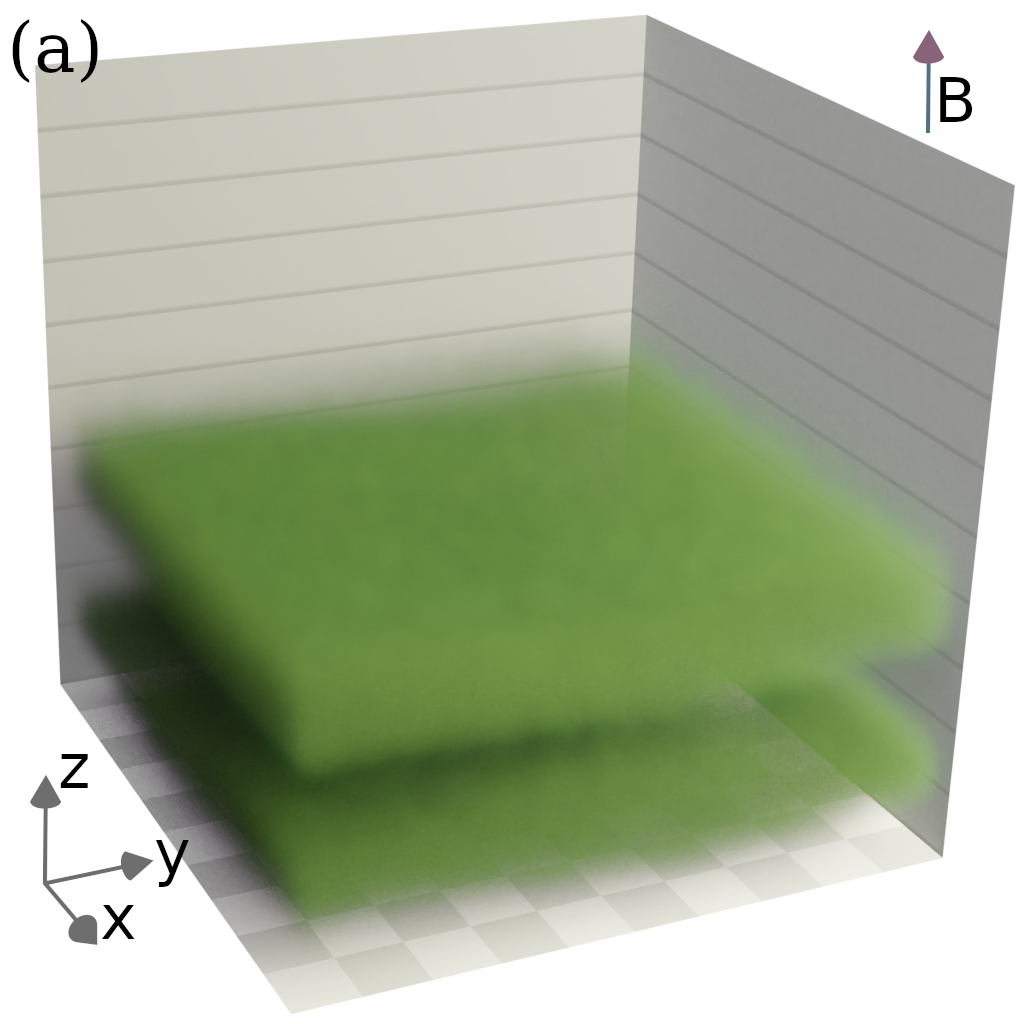}
  \includegraphics[width=0.45\columnwidth]{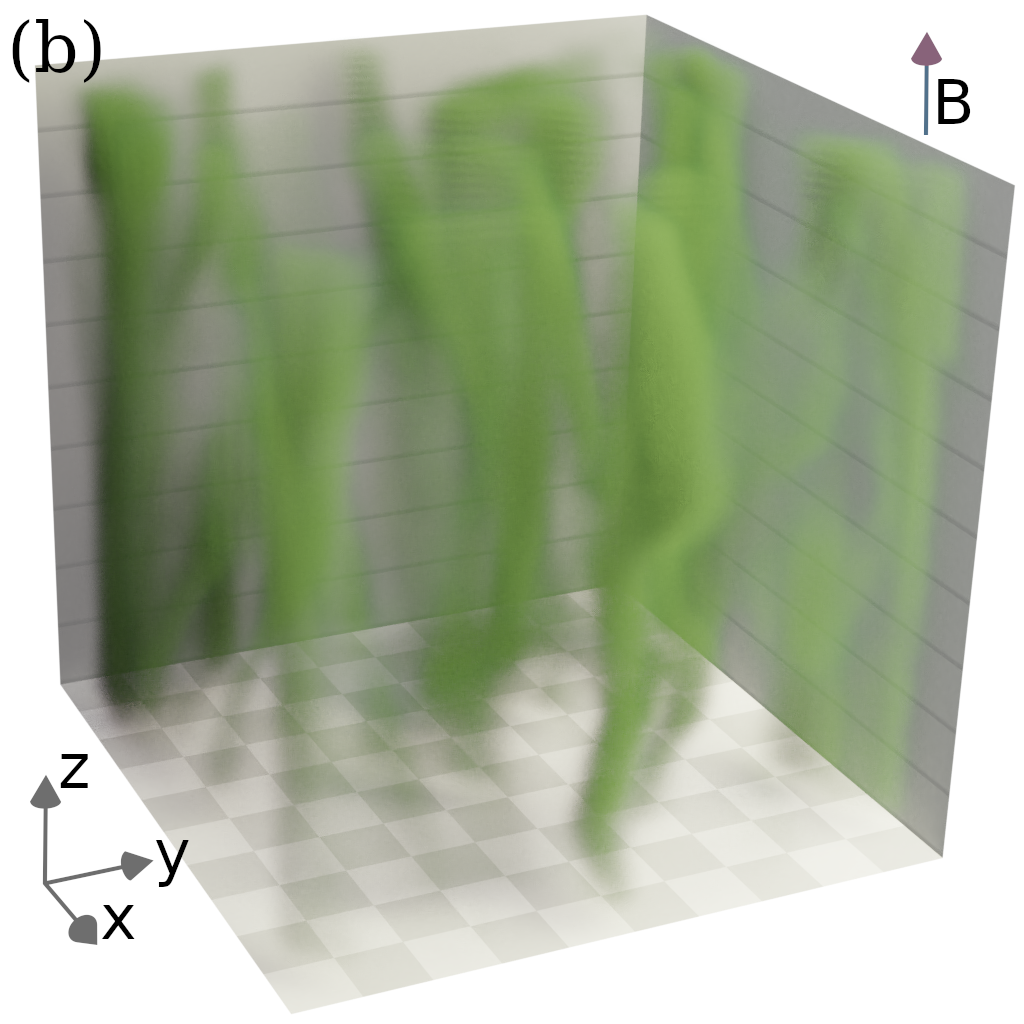}
  \caption{%
    3D volumetric rendering of the density field of a pusher (a) and puller (b) at \(t=1800\)  corresponding to the last time-step presented in panels (a) and (b) in Fig.~\ref{fig:hd_num_density}, respectively.
  }%
  \label{fig:hd_num_density_3d}
\end{figure}

\begin{figure}
  \centering
   \includegraphics[width=1.0\columnwidth]{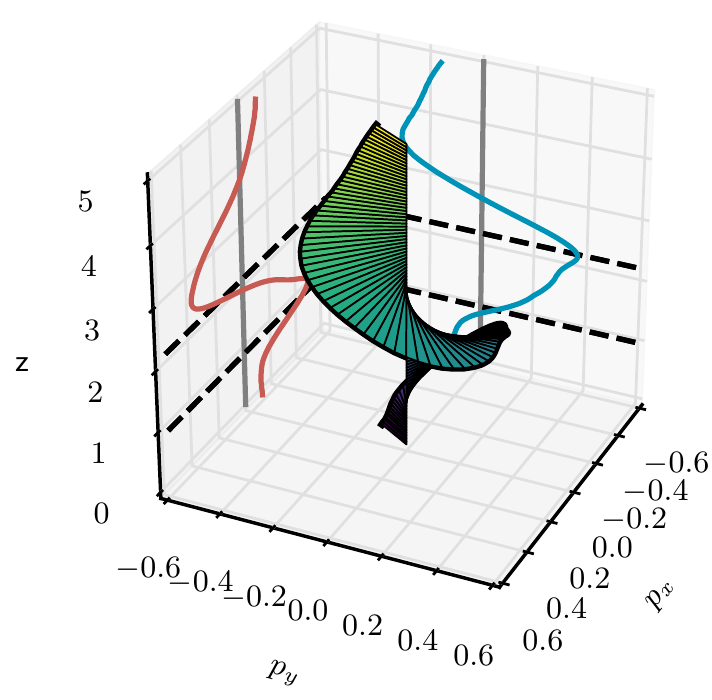}
  \caption{%
    3D representation of the radial component of the mean orientation \({\left<\vec{p}/\rho)\right>}_{x,y}\) revolving around the magnetic field axis in a helical twist pattern. The red and blue lines at the wall represent the respective projections in the are plotted, the black lines mark the positions of the density bands.
  }%
  \label{fig:hd_push1}
\end{figure}
  \begin{figure}
  \centering
   \includegraphics[width=\columnwidth]{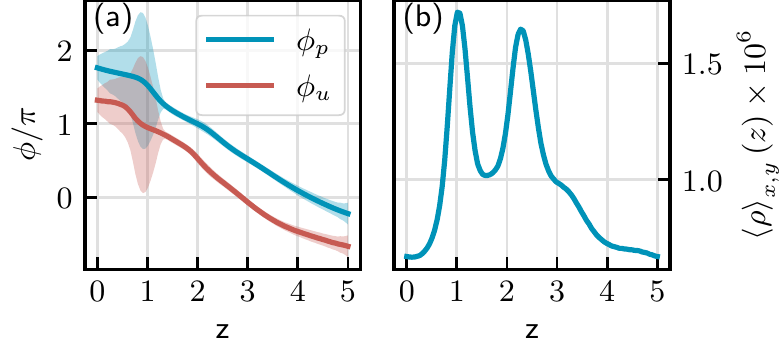}
  \caption{%
    (a) Mean azimuthal angles around the magnetic field axis for the orientation field \(\phi_p = \phi({\left<\vec{p}/\rho)\right>}_{x,y}\) and the flow field 
   \(\phi_u = \phi({\left<\vec{u}\right>}_{x,y})\).
  and (b) mean density along the magnetic field axis marking the position of two bands perpendicular to the magnetic field (compare with FIG.~\ref{fig:hd_num_density}(a)) for a representative snapshot of pushers with a dimensionless active stress \(\sigma_{\text{a}} = -30\) and alignment parameter \(\alpha_{\text{e}} = 4\) at $t=1800$. The magnetic field points along the \(z\)-axis.     }%
  \label{fig:hd_push2}
\end{figure}

The emerging picture from a 2D slice of instability snapshot, provides the ground for discussion of 3D patterns.  The 3D visualization of the density field shown in Fig.~\ref{fig:hd_num_density_3d}(a) is consistent with the picture drawn from a 2D slice. It clearly shows  that the pushers concentrate in band-like structures perpendicular to the magnetic field that migrate in the field direction $\hat{\vec{B}}\equiv \hat{\vec{z}}$. Now, if we plot  the  variation of the perpendicular component of the polarization field  averaged in the $x$-$y$ plane along the $z$-axis {\it i.e.} $\langle \vec{p}_{\bot}(z) \rangle_{x,y}=\langle(p_x(z),p_y(z)) \rangle_{x,y}$,  shown in Fig.~\ref{fig:hd_push1}, we observe a  helical-like evolution of $\langle \vec{p}_{\bot}(z) \rangle_{x,y}$. For clarity, each of the perpendicular components of polarization,
$\langle p_x(z) \rangle_{x,y}$ and $\langle p_y(z) \rangle_{x,y}$ are also shown in the $p_x$-$z$ and $p_y$-$z$ planes by red and blue lines, respectively. We note that each of the perpendicular components exhibit a bend-like instability. Therefore, the resultant  $\langle \vec{p}_{\bot}(z) \rangle_{x,y}$ can be interpreted as a superposition of phase shifted bend deformations of the orientation. The observed behavior is reminiscent of the bend-twist phase predicted for passive chiral or bent-shape liquid crystalline mesogens ~\cite{Meyer_1973,Dozov_2001} and observed experimentally in achiral molecules~\cite{Lavrentovich_2013}. Moreover,  in hydrodynamic theory  of vectorially ordered suspensions of  self-propelled particles it was predicted that the coupling between polar order and self-generated flow vorticity can lead to formation of bend-twist waves \cite{Simha_2002}.
In a bend-twist phase,  the polarization orientation vector draws an oblique helicoid, maintaining a constant oblique angle $0< \theta_0< \pi/2$  with the helix axis $z$: $\hat{\vec{p}}=(\sin \theta_0 \cos \phi, \sin \theta_0 \sin \phi, \cos \theta_0) $,  in which the azimuthal angle varies as $\phi=2 \pi z/\ell_{\text{p}} $ with  $\ell_{\text{p}}$ being the pitch of the helicoid.

To evaluate if the observed helicoidal pattern is associated with a bend-twist instability, we have extracted the values of the mean polar angle \( \theta_{\text{p}}(z) = \theta({\left<\vec{p}/\rho)\right>}_{x,y} \) and the azimuthal angle of  the mean  polarization  \( \phi_{\text{p}} (z)= \phi({\left<\vec{p}/\rho)\right>}_{x,y} \) averaged in the $x$-$y$ plane. We find that the mean polarization angle is almost independent of $z$, $\langle  \theta_{\text{p}}(z) \rangle \approx 0.475 \pi$, whereas 
   \( \phi_{\text{p}} \) shown  in Fig.~\ref{fig:hd_push2}(a) varies nearly linearly with $z$, apart from the regions close to the box boundaries. These results confirm that polar pushers in an alignment field are prevailed by a bend-twist instability. Moreover, we have also calculated  the azimuthal angle of  the flow field  averaged in the $x$-$y$ plane  \(\phi_{\text{u}}  = \phi({\left<\vec{u}\right>}_{x,y})\), in Fig.~\ref{fig:hd_push2}(a).   \(\phi_{\text{u}} \) also varies almost linearly with $z$, but shows a clear $\pi/2$ phase shift relative to  \(\phi_{\text{p}} \).  For comparison, we have also plotted the variation of   density averaged over the $x$-$y$ plane along the magnetic field axis in Fig.~\ref{fig:hd_push2}(b), which clearly shows the modulation of density as a result of band formation.  
\begin{figure}
  \centering
   \includegraphics[width=1.0\columnwidth]{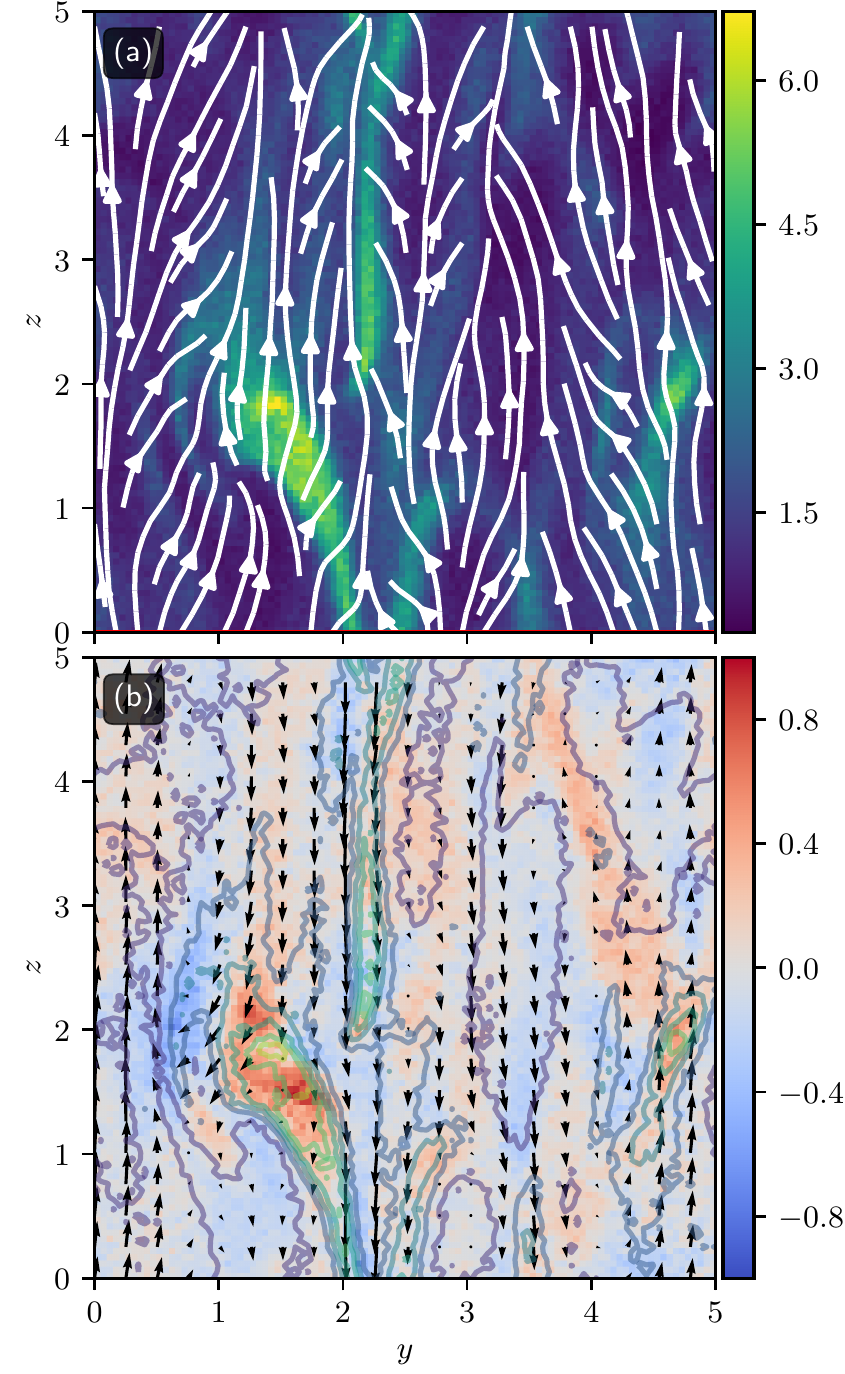}
  \caption{%
 (a)  Streamlines of the polarization field $(p_y,p_z)$ of a slice in the \(y\)-\(z\)-plane  at $x=0.75$,   demonstrating a characteristic splay patterns for a suspension of pullers with dimensionless active stress \(\sigma_{\text{a}} = 30\) and alignment parameter \(\alpha_{\text{e}} = 4\) in a magnetic field pointing in \(z\)-direction. The color encoded local density  \(\rho / V \times 10^6 \) is shown in the background. 
   (b): The corresponding flow field $(u_y,u_z)$ is represented as vector arrows, where the length of the vector  is weighted by its magnitude. The flow vorticity, leading to hydrodynamically induced particle rotation, is color encoded in the background. Red colors correspond to a vorticity vector pointing out of the plane (counter clockwise rotation).  Lines of constant  density are overlaid as  contour lines on top to guide the eyes. 
 }%
  \label{fig:hd_pull1_yz_slice}
\end{figure}

\begin{figure}
  \centering
   \includegraphics[width=1.0\columnwidth]{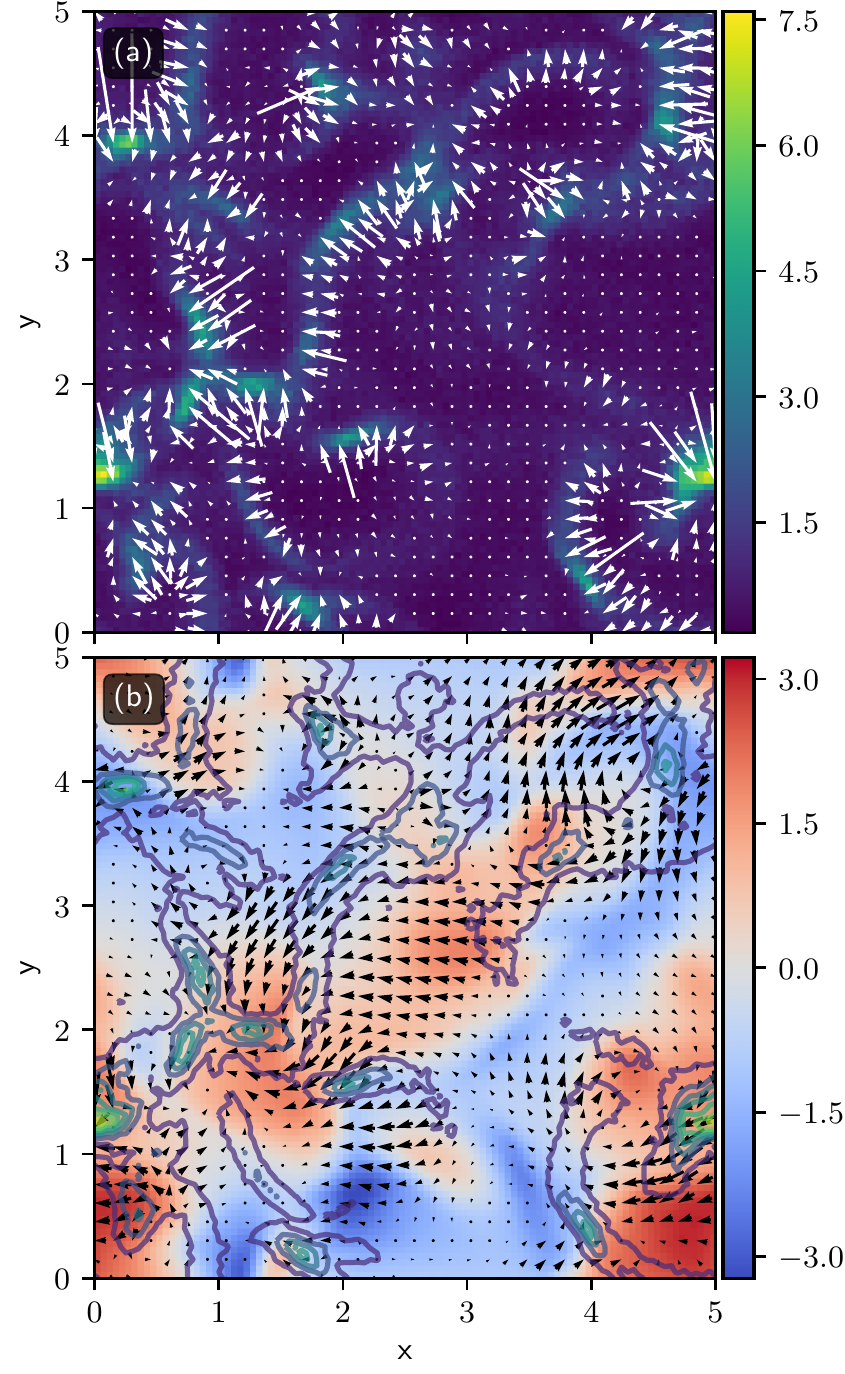}
  \caption{%
    (a):  Streamlines of the polarization field $(p_x,p_y)$ of a slice in the \(x\)-\(y\)-plane  at $z=0.75$,  where the size of arrows shows  the relative magnitude of polarization for a suspension of pullers with dimensionless active stress \(\sigma_{\text{a}} = 30\) and alignment parameter \(\alpha_{\text{e}} = 4\) for a magnetic field pointing in \(z\)-direction.  The color encoded local density \(\rho / V \times 10^6\) is superimposed in the background.  
   (b): The  flow field in the plane $(u_x,u_y)$ is represented as black arrows, where the length of the vector  is weighted by its magnitude. The \(z\)-component of flow field, $u_z$ is color encoded, where a red color denots a flow in the direction of the external magnetic field and a blue color opposed to it.  Lines of constant  density are overlaid as  contour lines on top to guide the eyes.
   }%
  \label{fig:hd_pull1_xy_slice}
\end{figure}

\subsubsection{Instability regime (b): moving pillars} 

The time series snapshots for pullers at the same alignment parameter \(\alpha_{\text{e}} = 4\) and activity strength \(\sigma_{\text{a}}=30\) are presented in Fig.~\ref{fig:hd_num_density} (b). At late stages, pullers tend to form dynamic pillar-like structures parallel to the external field axis. A 3D rendering of density field is shown in Fig.~\ref{fig:hd_num_density_3d}(b). 
Pullers exhibit a more complex density patterns relative to pushers with the same activity and magnetic field strength. This can be understood in light of 
 the linear stability analysis of homogeneous polar steady state, presented in panels (b1) and (b2) of Fig.~\ref{fig:hd_growthrate_k}, which predicts the predominance of long wavelength instabilities with wavevectors both parallel and  perpendicular to the magnetic field. Although,  our non-linear dynamics simulations  display some density undulations in the direction parallel to the magnetic field, we observe that pattern formation is primarily prevailed by the perpendicular perturbations as predicted by the linear stability of  the moment equations, Eqs.~\eqref{eq:hd_density_time} and \eqref{eq:hd_polarization_time}.

 Next, we examine the corresponding polarization and flow fields on a 2D slice of the sample in the $y$-$z$ plane.  Fig.~\ref{fig:hd_pull1_yz_slice}(a)  displays the polarization  field  superimposed by the density field at a late stage $t=1800$ and the corresponding flow and vorticity fields are presented in Fig.~\ref{fig:hd_pull1_yz_slice} (b).  We observe that the polarization streamlines significantly deviate from straight lines. The distortions of polarization field can be understood in terms of splay instabilities in the language of liquid crystals~\cite{LC_Prost}, see also the appendix  for an idealized description of bend and splay distortions. 
 
 Splay deformations, similar to bend fluctuations,  consist of small polarization perturbations which are perpendicular to \(\vec{p}_0 \parallel \hat{\vec{B}} \equiv \hat{\vec{z}} \)
  but in this case  their magnitude  is modulated in directions perpendicular to it, {\it i.~e.}, $\vec{p}_p=\tilde{\vec{p}}_{\bot}  \exp( \I  \vec{k}_{\bot} \cdot \vec{x}_{\bot} )$, where $\vec{k}_{\bot}\equiv (k_x,k_y,0)$ and $\vec{x}_{\bot}\equiv (x,y,0)$~\cite{LC_Prost,ramaswamy_active-filament_2007}.  Again based on  Eq.~\eqref{eq:hd_density_time} for the density moment, splay fluctuations increase the density  in volumes of negative divergence \(\nabla \cdot \vec{p} < 0\) and generate alternating pillar-like flow regions along \(\hat{\vec{B}}\) as shown in Fig.~\ref{fig:hd_pull1_yz_slice}(a); see also the appendix for illustration of an idealized splay distortion.   Splay distortions also generate a  position-dependent active force density $\vec{f} \approx  \sigma_{\text{a}} \nabla \cdot \left( \vec{p} \vec{p} - \frac{1}{3} \mathds{1} \right)$ in the fluid that result in alternating flow layers parallel and anti-parallel to the magnetic field, see  Fig.~\ref{fig:hd_pull1_yz_slice}(b). The  vorticity field   also become heterogenous and induces a hydrodynamic torque \(\vec{M}_{\text{HD}} \propto \frac{1}{2} \nabla \times \vec{u} \). This torque rotates the swimmers further away from the magnetic field axis  and renders  a uniform homogenous polar phase unstable. The more concentrated regions of pullers, where \(\nabla \cdot \vec{p} < 0\) coincide with regions  carrying a flow anti-parallel to \(\hat{\vec{B}}\) and high self-generated flow vorticity. They result in a net  convection anti-parallel to the magnetic field and  reduce the mean transport speed~\cite{Koessel_2019}. Furthermore, we have also shown  the polarization  field  superimposed by the density field   and the corresponding flow and vorticity fields  of a $x$-$y$ slice perpendicular to the magnetic field at $z=0.75$ in Fig.~\ref{fig:hd_pull1_xy_slice}.   We note that the perpendicular component of polarization is largest in denser regions. Consistent with our picture from a $y$-$z$ slice, the regions of large vorticity are correlated with the concentrated regions of swimmers.

 \subsubsection{Instability regimes (c) and (d):  finite-sized concentrated regions }
 \begin{figure}
  \centering
   \includegraphics[width=1.0\columnwidth]{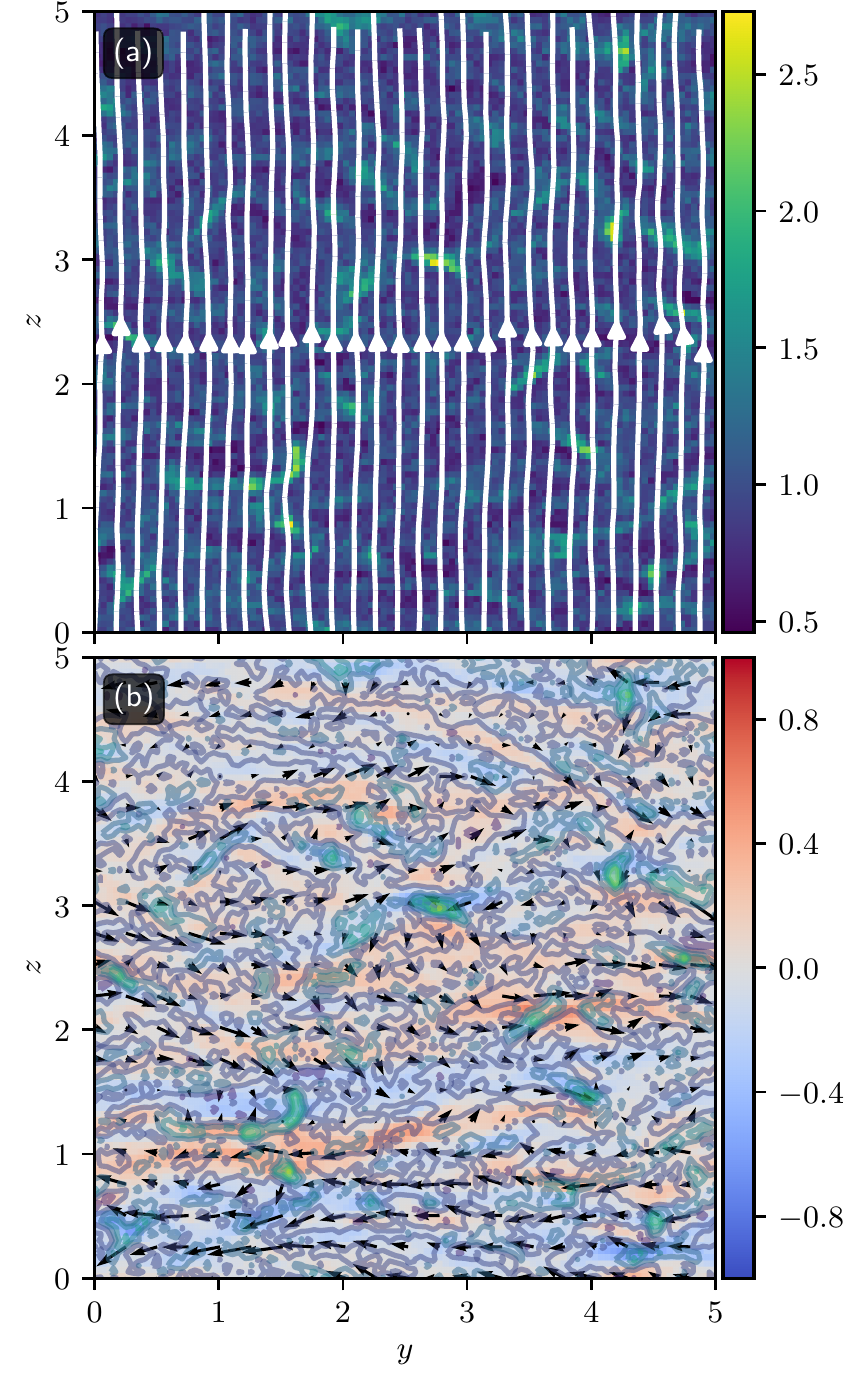}
  \caption{%
 (a)  Streamlines of the polarization field $(p_y,p_z)$ of a slice in the \(y\)-\(z\)-plane  at $x=0.75$,   for a suspension of pullers with dimensionless active stress \(\sigma_{\text{a}} = -40\) and alignment parameter \(\alpha_{\text{e}} = 19\) in a magnetic field pointing in \(z\)-direction. The color encoded local density  \(\rho / V \times 10^6 \) is shown in the background. 
   (b): The corresponding flow field $(u_y,u_z)$ is represented as vector arrows, where the length of the vector  is weighted by its magnitude. The flow vorticity, leading to hydrodynamically induced particle rotation, is color encoded in the background. Red colors correspond to a vorticity vector pointing out of the plane (counter clockwise rotation).  Lines of constant  density are overlaid as  contour lines on top to guide the eyes.    }%
  \label{fig:hd_push_c_yz_slice}
\end{figure}

\begin{figure}
  \centering
   \includegraphics[width=1.0\columnwidth]{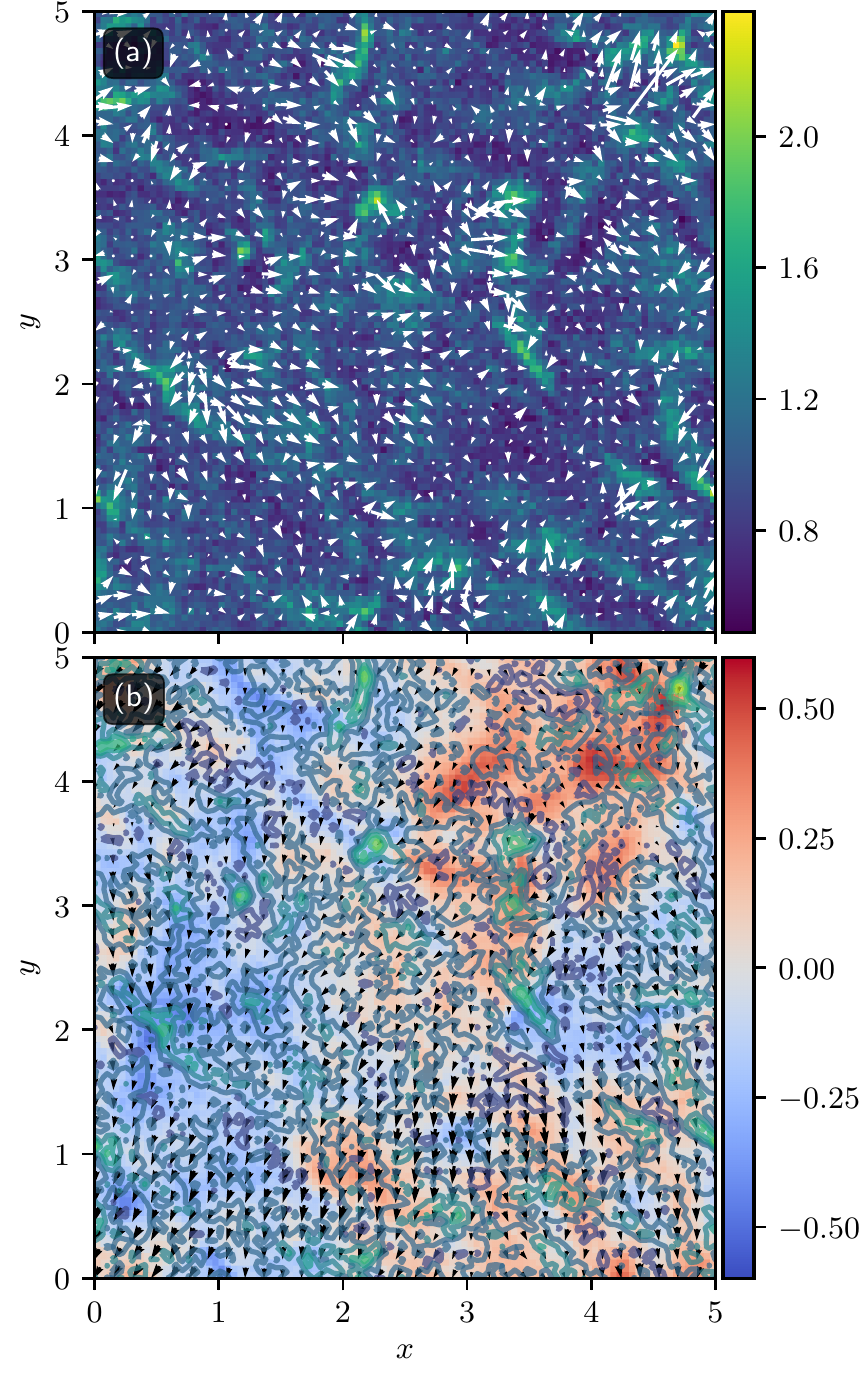}
  \caption{%
      (a):  Streamlines of the polarization field $(p_x,p_y)$ of a slice in the \(x\)-\(y\)-plane  at $z=0.75$,  where the size of arrows shows  the relative magnitude of polarization for a suspension of pushers with dimensionless active stress \(\sigma_{\text{a}} = -40\) and alignment parameter \(\alpha_{\text{e}} = 20\) for a magnetic field pointing in \(z\)-direction.  The color encoded local density \(\rho / V \times 10^6\) is superimposed in the background.  
   (b): The  flow field in the plane $(u_x,u_y)$ is represented as black arrows, where the length of the vector  is weighted by its magnitude. The \(z\)-component of flow field, $u_z$, is color encoded with a red color denoting a flow in the direction of the external magnetic field and a blue color opposed to it.  Lines of constant  density are overlaid as  contour lines on top to guide the eyes.  }%
  \label{fig:hd_push_c_xy_slice}
\end{figure}

 \begin{figure}
  \centering
   \includegraphics[width=1.0\columnwidth]{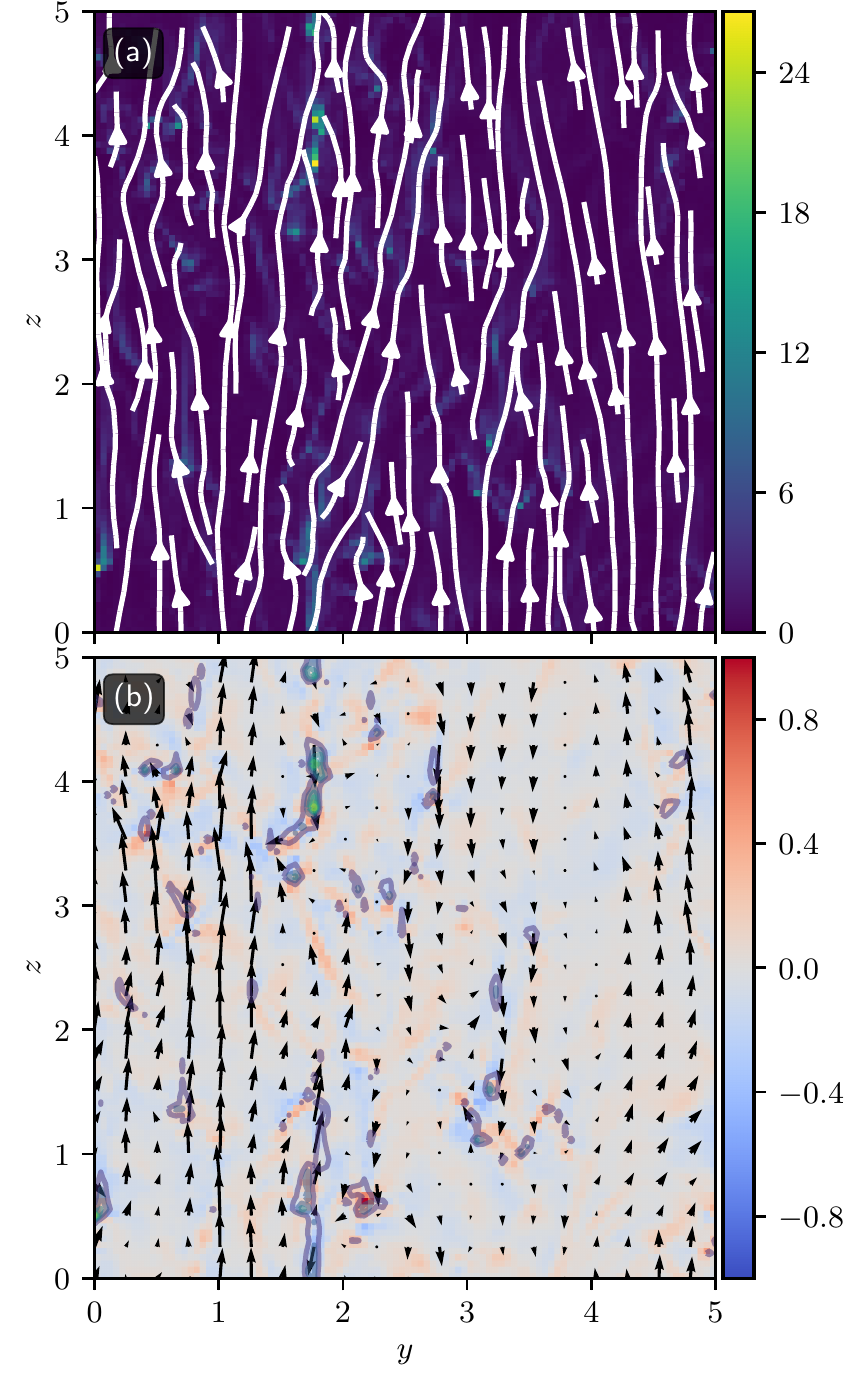}
  \caption{%
 (a)  Streamlines of the polarization field $(p_y,p_z)$ of a slice in the \(y\)-\(z\)-plane  at $x=0.75$,  exhibiting weak   splay distortions for a suspension of pullers with dimensionless active stress \(\sigma_{\text{a}} = 40\) and alignment parameter \(\alpha_{\text{e}} = 19\) in a magnetic field pointing in \(z\)-direction. The color encoded local density  \(\rho / V \times 10^6 \) is shown in the background. 
   (b): The corresponding flow field $(u_y,u_z)$ is represented as vector arrows, where the length of the vector  is weighted by its magnitude. The flow vorticity, leading to hydrodynamically induced particle rotation, is color encoded in the background. Red colors correspond to a vorticity vector pointing out of the plane (counter clockwise rotation).  Lines of constant  density are overlaid as  contour lines on top to guide the eyes.   }%
  \label{fig:hd_pull_d_yz_slice}
\end{figure}

\begin{figure}
  \centering
   \includegraphics[width=1.0\columnwidth]{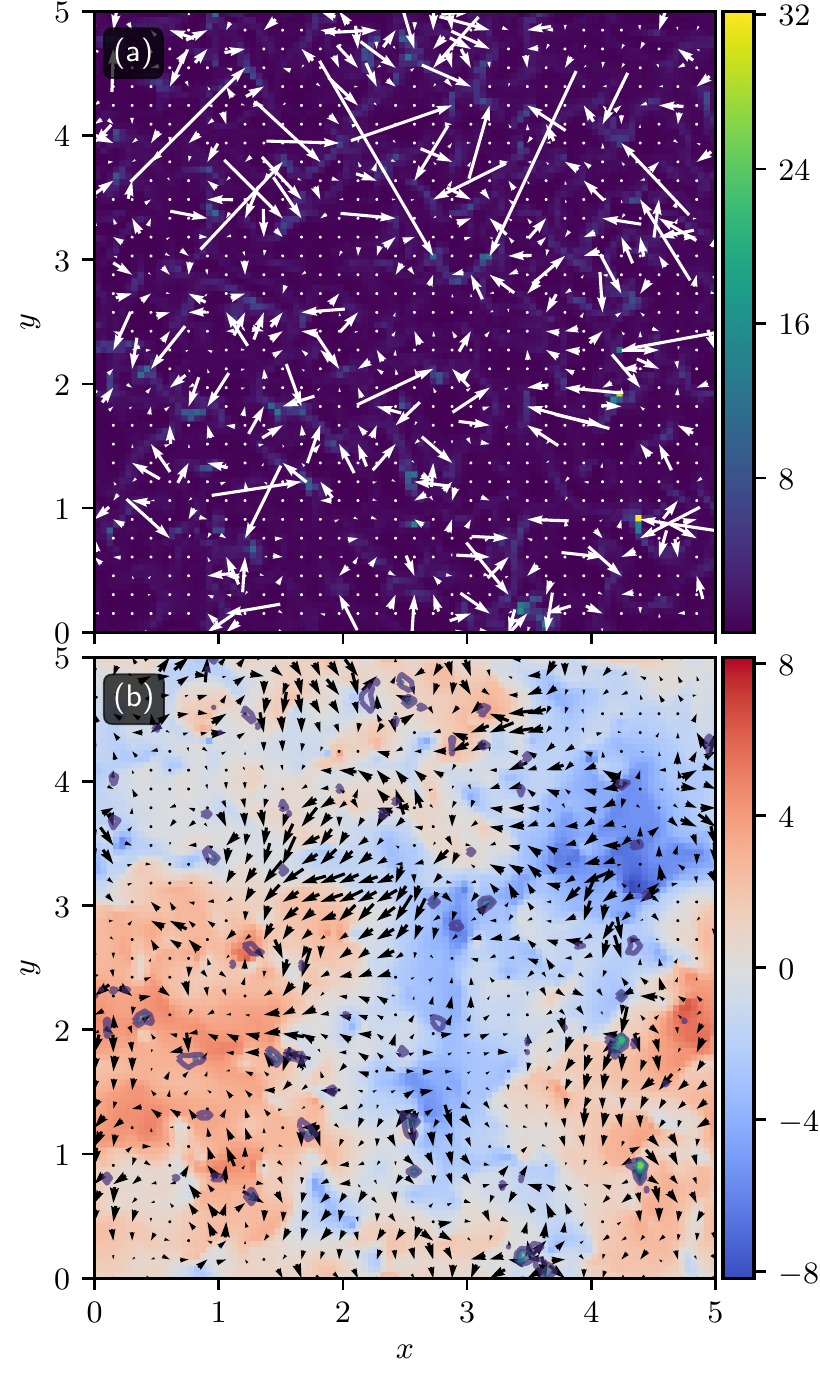}
  \caption{%
   (a):  Streamlines of the polarization field $(p_x,p_y)$ of a slice in the \(x\)-\(y\)-plane  at $z=0.75$,  where the size of arrows shows  the relative magnitude of polarization for a suspension of pullers with dimensionless active stress \(\sigma_{\text{a}} = 40\) and alignment parameter \(\alpha_{\text{e}} = 20\) for a magnetic field pointing in \(z\)-direction.  The color encoded local density \(\rho / V \times 10^6\) is superimposed in the background.  
   (b): The  flow field in the plane $(u_x,u_y)$ is represented as black arrows, where the length of the vector  is weighted by its magnitude.The \(z\)-component of flow field, $u_z$,  is color encoded with a red color denoting a flow in the direction of the external magnetic field and a blue color opposed to it.  Lines of constant  density are overlaid as  contour lines on top to guide the eyes.
  }%
  \label{fig:hd_pull_d_xy_slice}
\end{figure}

The snapshots in  Fig.~\ref{fig:hd_num_density} (c) and (d) show the evolution of density profiles of pushers with  \(\alpha_{\text{e}} =19\) and \(|\sigma_{\text{a}}|=40\).
 They correspond to the  panels (c) and (d) of Fig.~\ref{fig:hd_growthrate_k}, where the linear stability analysis  of homogeneous polar steady state predicts predominance of perturbations with a finite wavelength. In both cases, the length scale associated with concentrated regions is finite, $k_{\text{max}} \approx 2.5$, and smaller than the box size as opposed to cases (a) and (b), where the maximum growth rate occurs at the limit $k \to 0$.   However, the  angle of wavevector relative to the magnetic field $ \Theta_B$, for which the the maximum growth rate occurs is different for pushers $ \Theta_B^{\text{max}}\approx 30 $ and pullers $ \Theta_B^{\text{max}} \approx 60 $. 
 In both cases, we see fluctuating concentrated regions which on the average migrate in the direction of magnetic field suggesting that some kind of dynamical aggregates are formed.  Consistent with the stability analysis prediction, the morphology of the aggregates  are different for pushers and pullers with the identical activity and magnetic field strengths. To gain more insight into similarities and differences between pushers and pullers, we look into the polarization and flow fields in each case.
  
    For pushers in the case (c), concentrated regions form bands with a finite length and a wide range of angles relative to the magnetic field. These patterns are  distinct from  those of pushers in the case (a) of Fig.~\ref{fig:hd_num_density}, where bands  are perpendicular to the magnetic field and expand the whole lateral dimension of the box; see Fig.~\ref{fig:hd_num_density_3d}(a). Looking into the polarization field in a $y$-$z$ slice shown in Fig.~\ref{fig:hd_push_c_yz_slice}(a), we observe very weak deviations from a uniform polarization,  whereas  the density is notably heterogenous. The generated flow field and its associated vorticity are shown in Fig.~\ref{fig:hd_push_c_yz_slice}(b) and they are  weaker than the flow and vorticity created in the case (a) presented in Fig.~\ref{fig:hd_push_yz_slice}(b), which is predominated by the bend-twist instability.  Examining the  polarization field superimposed by density  in a $x$-$y$ slice perpendicular to the magnetic field  shown in Fig.~\ref{fig:hd_push_c_xy_slice}(a), we find that the deviations of polarization field from the  \( \hat{\vec{B}} \equiv \hat{\vec{z}} \) occur at  concentrated regions. In other words, the finite wavelength instabilities are dominantly density driven  and the polarization distortions merely stem from density perturbations.  In the language of linear stability analysis, the predominant mode of perturbation is given by the mode shape $ \psi_0^0(\vec{k}_{\text{max}}) e^{\I \vec{k}_{\text{max}} \cdot\vec{x} + \lambda t} \propto  \rho_p(\vec{k}_{\text{max}}, \vec{x}, t) $, where $\vec{k}_{\text{max}}$ corresponds to the wavevector with the largest growth rate shown in Fig.~\ref{fig:hd_growthrate_k}(c). Likewise, from Fig.~\ref{fig:hd_push_c_xy_slice}(b), we note that the self-generated flow velocity  component perpendicular to the magnetic field $(u_x,u_y) $ is rather weak and it only becomes considerable in concentrated regions. Unlike the case (a), the flow field has an appreciable component along the magnetic field (parallel or anti-parallel) as encoded by red and blue colors in Fig.~\ref{fig:hd_push_c_xy_slice}(b).

 For the pullers in the stability regime (d), the long-time density pattern resembles that of  pullers with moderate strengths of the activity and magnetic field in case (b). However, concentrated regions consist of   of finite-sized pillar-like aggregates  in contrast to the case (b), where pillar-like dense regions span the whole box dimension in the field direction, verifying the predominance of smaller wavelength density fluctuations. Moreover, finite-sized concentrated regions are on average not parallel to the magnetic field and have a wider orientation distribution. Inspecting the polarization field superimposed by the density field in a $y$-$z$ slice perpendicular to the magnetic field  shown in Fig.~\ref{fig:hd_pull_d_yz_slice}(a), we notice that  polarization field displays some splay deformations.  However, its distortions are weaker in comparison to the case (b).  Likewise, the  self-generated flow field is very similar to the case (b) and we observe a notable vorticity field in  concentrated regions. Looking into the polarization and velocity fields in a $x$-$y$ slice perpendicular to 
\( \hat{\vec{B}} \equiv \hat{\vec{z}} \), we find that similar to case (c), the deviations from a uniform polarization occurs at concentrated regions, which lead  to a very heterogenous flow field  as shown in Fig.~\ref{fig:hd_pull_d_yz_slice}(b). Despite the similarities of polarization and flow field with the case (b), the  predominant mode of perturbation in the linear stability is  the density  mode similar to the case (c). It  is given by  $ \psi_0^0(\vec{k}_{\text{max}}) e^{\I \vec{k}_{\text{max}} \cdot\vec{x} + \lambda t} \propto  \rho_p(\vec{k}_{\text{max}}, \vec{x}, t) $, where $\vec{k}_{\text{max}}$ corresponds to the wavevector with the largest growth rate shown in Fig.~\ref{fig:hd_growthrate_k}(d). 
 
 
\section{DISCUSSION AND CONCLUDING REMARKS}%
\label{sec:conclusion}
We have presented a continuum kinetic model for  active suspensions of weakly magnetic spherical particles  in an external field. 
 The model is based on first principles, namely, a conservation equation for the particle configuration distribution in an alignment field, coupled to the Stokes equation for the fluid flow which incorporates stress contributions steming from activity and alignment torque.  It is applicable to moderately dilute suspensions of magnetotactic bacteria or artificial magnetic microswimmers with a small magnetic moment and focuses on the interplay between hydrodynamic interactions arising from self-generated flow and external alignment torque. 
 We investigated the nature of hydrodynamic instabilities and emergent pattern formation 
 by combining linear stability analysis and the full numerical solution of kinetic model equations.

 We first performed a linear stability analysis of steady state solution of the model, which corresponds to a homogenous polar distribution function $\psi_0$. The stability analysis of  steady state as a function of activity and magnetic field strengths reveals that a uniformly polarized suspension  becomes unstable for moderate magnetic field and sufficiently large activity strengths for both pushers and pullers.  Based on the dispersion relation of the maximum growth rate, we have drawn a  non-equilibrium phase diagram as  presented in Fig.~\ref{fig:hd_stab_dia}.  We recognize four distinct  instability regimes.  For moderate activity and field strengths, the long wavelength instabilities dominate  both pusher  and puller suspensions. However, the nature of instabilities are different for the two types of swimmers. Pushers are dominated by wave perturbations parallel to the field, whereas pullers are unstable with respect to both parallel and perpendicular wave perturbations. For stronger activities and magnetic fields, the wavevector with the largest growth rate has a finite wavelength and its angle with the field  differs for pushers and pullers with the same activity and field strengths. These  instability regimes are driven by  density fluctuations as opposed to long wavelength instabilities which are driven by the orientational fluctuations.
 Increasing the magnetic field strength further, the  alignment torque is strong enough to overcome  the hydrodynamically induced torque. As a consequence,  the homogenous polar state becomes stable again and we observe a \emph{reentrant hydrodynamic stability}.  
 
   Next,  we obtained the  dynamical equations for the first three orientational moments, \emph{i.e.}, density, polarization and nematic field, imposing suitable closure approximations. Moment equations, although less accurate,  provide us with new insights into the nature of instabilities. 
 As can be seen from the moment equations, Eqs.~\eqref{eq:hd_density_time}, \eqref{eq:hd_polarization_time} and \eqref{eq:hd_nematic_time},  density, polarization and nematic fields are coupled to each other. This implies that any heterogeneity in one of them generates a heterogeneity in the other fields leading to a feedback loop until a new dynamical equilibrium is reached. 
  Linear stability analysis of moment equations for uniform density and polarization fields predicts the predominance of  long wavelength  instabilities with wavevectors parallel to the alignment field  for pushers and wavevectors perpendicular for pullers  at moderate magnetic fields. Based on these results, we deduce that pushers are prevailed by \emph{bend} deformations, whereas pullers are predominated by \emph{splay} distortions.
 These findings are in agreement the linear stability analysis of the steady distribution function $\psi_0$  for a large region of stability diagram, where long wavelength instabilities  prevail the system,  although perturbation of $\psi_0$, equivalent to considering the full hierarchy of moments, predicts the predominance of both long wavelength parallel and perpendicular perturbations for pullers. Moreover, the coarse-grained approximate moment equations do not capture the  finite wavelength instability regimes at higher magnetic fields and activity strengths.

 To evaluate the accuracy of predictions of the linear stability analysis, we investigated the numerical solution of kinetic model equations. Numerical simulations show very good agreement with predictions of linear stability analysis for the borderlines of instability. Although linearly unstable modes do not capture the full non-linear dynamics, many aspects of the dynamics observed  in simulations can be understood in the light of the stability analysis.   According to Fig.~\ref{fig:hd_stab_dia}, for a large region in the parameter space the most unstable mode for pushers is parallel to the external field, whereas for pullers both parallel and perpendicular unstable modes dominate the system. Simulations show that indeed  in a large part of the unstable region   predominant modes of  instability  for pushers are \emph{bend-twist} distortions of the polarization field. For the pullers, \emph{splay} deformations prevail the pattern formation  suggesting that  the perpendicular perturbation is the predominant mode of deformation.  As a consequence, we observe distinct patterns for the two kinds of swimmers: traveling bands perpendicular to the magnetic field for pushers and pillar-like concentrated regions parallel to the field for pullers.   
   As discussed in our prior work~\cite{Koessel_2019}, the deflections of polarization field lead to a reduction of the average polarization and mean transport speed. 
In the regions of  stability diagram of Fig.~\ref{fig:hd_stab_dia}, where  the maximum growth rate occurs at finite wavelengths, we observe finite-sized  concentrated regions  suggesting formation of dynamical aggregates in external field. However, the morphology of these regions is different for pushers and pullers in agreement with predictions of linear stability analysis.   
 
 We conclude by pointing out a few limitations of the present model  and future directions. Our results are obtained in the limit of negligible magnetic interactions and only consider the interaction of a single particle
with a mean-field flow.  This limits the validity of our model to moderately dilute suspensions to   magnetic swimmers with weak dipole moments such as magnetotactic bacteria. Nevertheless, we believe that the present model captures most salient features of interplay between hydrodynamic interactions and external field in not so concentrated active suspensions. For instance,  band formation observed for pushers are in agreement with  experimental findings of  magnetotactic bacteria at moderate field strengths $B \sim 3 $ mT~\cite{spormann_unusual_1987}. 
 For synthetic magnetic microswimmers with larger magnetic dipole moments or dense suspensions, the magnetic dipolar interactions alone can lead to clustering instabilities~\cite{meng_clustering_2018} and the interplay between long-range magnetic and hydrodynamic interactions on development of instabilities merits further investigations.
Moreover,   the role of swimmer-swimmer correlations~\cite{CorrPRL2017}, and near-field hydrodynamic interactions in more concentrated suspensions remains an open question.
Finally, our results show that a sufficiently strong  alignment field can overcome hydrodynamic instabilities calling for further exploration of controlling the collective behavior and transport of active matter in various external fields.

\begin{acknowledgments}
We thank Oleg Lavrentovich for fruitful discussions regarding the bend-twist phases. 
We acknowledge the financial support from the German Research Foundation (http://www.dfg.de) within SFB TRR 146 (https://trr146.de). The simulations were performed using the MOGON II computing cluster. This research was supported in part by the National Science Foundation under Grant No. NSF PHY-1748958.          
\end{acknowledgments}

\appendix
\section{Idealized bend and splay instabilities}%
\label{sec:hd_bend_and_splay_ideal}
To illustrate better the underlying mechanism of the pattern formation for pushers and pullers in an external field, we present here the idealized 2D \emph{bend}  and \emph{splay} deformations. Hence, we restrict the discussion to a plane with components parallel and perpendicular to the external field, \(x_{\parallel}\equiv \hat{\vec{z}}\) and \(x_{\perp}\), respectively. We   approximate the polarization field  of a \emph{bend}-deformation by
\begin{equation*}
  \vec{p}_{\mathrm{bend}}(x_{\parallel}) = \left( \tilde{p}_{\perp} \cos (k x_{\parallel}),\sqrt{1 - \tilde{p}_{\perp}^2 \cos ^2(k x_{\parallel})} \right),
\end{equation*}
assuming that the magnitude of polarization is constant everywhere and its perpendicular component varies as a cosine function of  a wavenumber \(k\) along the field direction with an amplitude \(\tilde{p}_{\perp}\). Likewise, we approximate a splay deformation is  by a polarization field of constant magnitude where the  perpendicular component varies as a cosine function of  amplitude \(\tilde{p}_{\perp}\) and   wavenumber \(k\)  modulated in the direction perpendicular to the field:
\begin{equation*}
  \vec{p}_{\mathrm{splay}}(x_{\perp}) =\left( \tilde{p}_{\perp} \cos (k x_{\perp}),\sqrt{1-\tilde{p}_{\perp}^2 \cos^2(k x_{\perp})} \right).
\end{equation*}
\begin{figure}[h]
  \centering
   \includegraphics[width=0.99\linewidth]{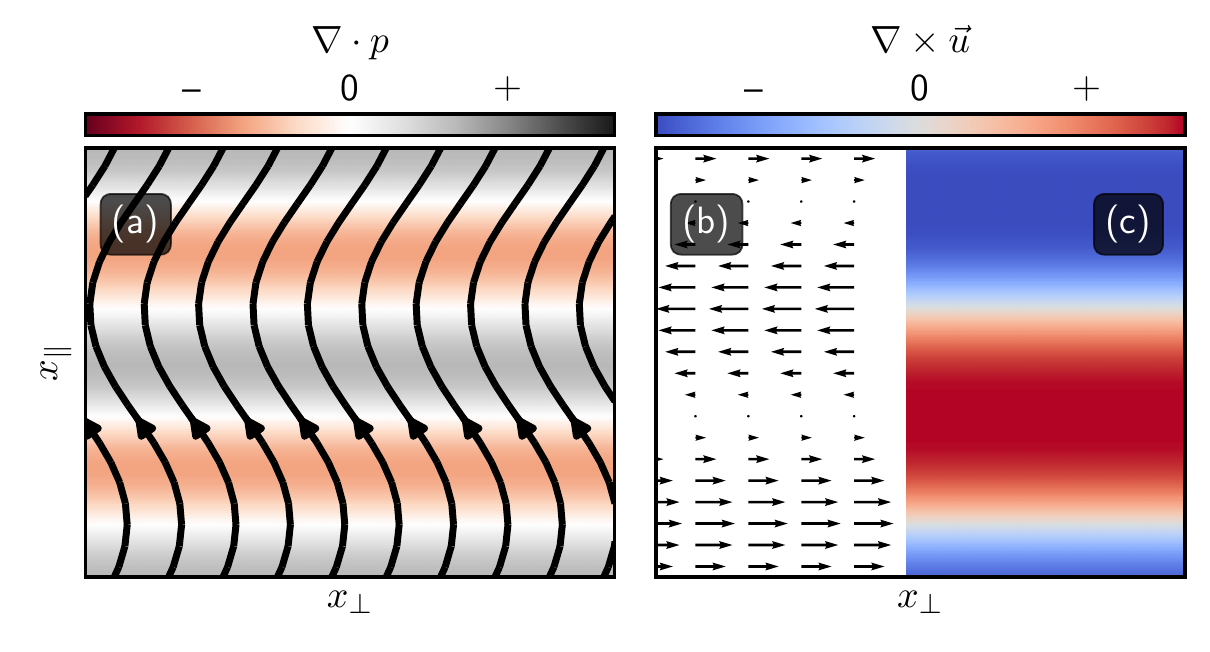}
  \caption{%
    An idealized (static) \emph{bend} deformation for polarized pushers (in \(x_{\parallel}\)-direction). (a) shows the streamlines of the polarization field \(\vec{p}\) with its divergence \(\nabla \cdot \vec{p}\) color coded in the background. Swimmers accumulate eventually in volumes of negative divergence (red). (b) Created flow field perpendicular to polarization axis. (c) The resulting  flow vorticity field  enhances the \emph{bend} perturbation further (red color CCW rotation, blue color CW).
  }%
  \label{fig:hd_bend_idealized}
\end{figure}

\begin{figure}[h]
  \centering
   \includegraphics[width=0.8\linewidth]{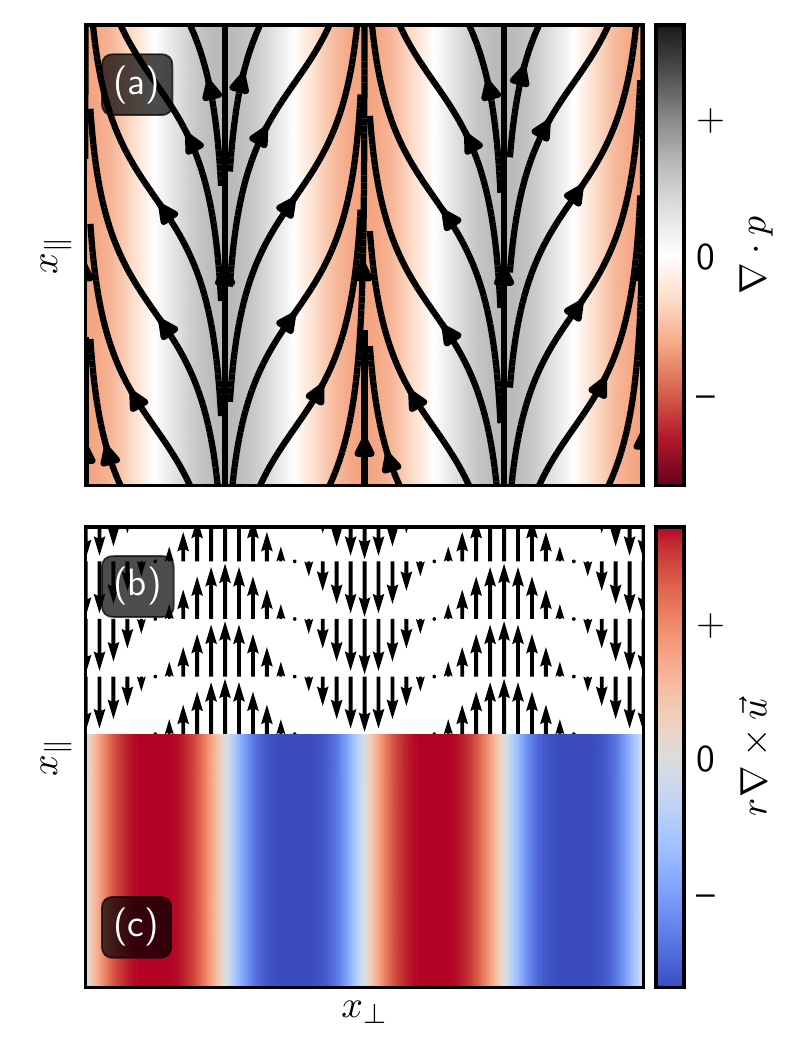}
  \caption{%
    An idealized (static) \emph{splay} deformation for polarized pullers (in \(x_{\parallel}\)-direction) analogous to Fig.~\ref{fig:hd_bend_idealized}.
  }%
  \label{fig:hd_splay_idealized}
\end{figure}
\begin{figure}
  \centering
   \includegraphics[width=0.95\linewidth]{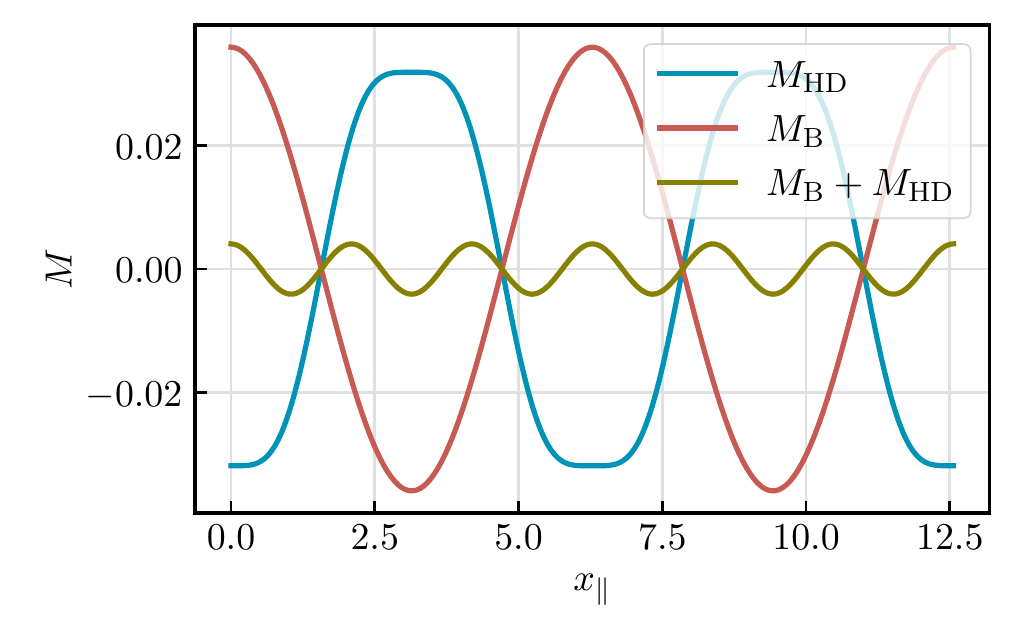}
  \caption{%
    Illustration of the competition between a (virtual) hydrodynamic torque \(M_{\text{HD}} \propto \frac{1}{2} \|\nabla \times \vec{u}\|\) and an external magnetic torque \(M_{\text{B}} \propto \alpha_{\text{e}} \| \vec{n} \times \hat{\vec{B}}_{\text{e}}\|\) for \emph{bend} and \emph{splay} instabilities in arbitrary units. The instabilities cause the flow vorticity to grow while rotating the particles, until the rotation is compensated by the external torque (amber).
  }%
  \label{fig:hd_bend_torque}
\end{figure}
The  active force density  resulting from such a polarization field can then be estimated as
\begin{equation*}
  \vec{f}_\text{a} \approx \sigma_{\text{a}} \nabla \cdot \left( \vec{p} \vec{p} - \frac{1}{3} \mathds{1} \right).
\end{equation*}
 For simplicity, we assume \(\sigma_{\text{a}}=\pm 1\). Given the force density field, the Stokes equation
\begin{equation*}
  \Delta \vec{u} - \nabla p + \vec{f}_\text{a} = 0
\end{equation*}
can be solved using the Oseen tensor and a spectral method. The flow velocity is perpendicular to the external field for pushers \(\vec{u} = u_{\perp}(x_{\parallel}) \hat{\vec{x}}_{\perp}\) and parallel to the field for pullers \(\vec{u} = u_{\parallel}(x_{\perp}) \hat{\vec{x}}_{\parallel}\). To very good approximation, it can be described by
\begin{equation*}
  u_{\{\perp, \parallel\}}(\{x_{\parallel}, x_{\perp}\}) \approx c_1 \sin \{x_{\parallel}, x_{\perp}\} + c_2 \sin 3 \{x_{\parallel}, x_{\perp}\},
\end{equation*}
in which the braces \( \{ \} \) evaluate to the first entry for pushers, the second entry for pullers, and \(c_1, c_2 \in \mathbb{R}\) are some numerical prefactors. 

The polarization field and the associated flow and vorticity fields  of pushers with \emph{bend} deformations and pullers with \emph{splay} distortions in the alignment field are  shown in Fig.~\ref{fig:hd_bend_idealized} and Fig.~\ref{fig:hd_splay_idealized}, respectively. The subplots  (a)  shows  the streamlines of the polarization field \(\vec{p}\) with its divergence color encoded in the background. The swimmers concentrate in the red areas where the divergence is negative. As a consequent, pushers form band-like regions  perpendicular to the field, whereas  pullers concentrate in  pillar-like dense regions parallel to the field. The subplots (b) depict a vector plot of the self-generated flow fields as a result of the bend and splay deformations of the polarization fields of pushers and pullers, respectively. In both cases, we observe alternating flow layers modulated in the  direction perpendicular to field. For the pushers, the  flow velocities  with alternating directions are perpendicular to the alignment field, while for pullers the the alternating flow velocities are parallel and anti-parallel to the field. In subplots (c),  the flow fields' vorticity fields \(\nabla \times \vec{u}\) are shown, where the red color encodes a counter clockwise rotation and the blue color a clockwise rotation. We observe the alternating clockwise and anticlockwise vorticity fields  are also modulated in directions  parallel perpendicular to the field for pushers and pullers, respectively. 
 According to the Faxen's law,  the flow vorticity  induces effectively a hydrodynamic torque $  \vec{M}_{\text{HD}} \propto \frac{1}{2} \nabla \times \vec{u}$, which rotates the particles away from the magnetic field direction and competes with the magnetic torque. A homogenous polar steady state becomes unstable due to these competing torques.
%

In the idealized case, both torques are easy to calculate and are plotted in arbitrary units in Fig.~\ref{fig:hd_bend_torque} for a \emph{bend}-deformation (it is qualitatively the similar for \emph{splay}-deformations). The mean, dimensionless external torque, given by \(\vec{M}_{\mathrm{B}} = \alpha_{\text{e}} \vec{p} \times \hat{\vec{B}}\) nearly fully balances the \(M_{\text{HD}}\) hindering further growth of bend deformation. As a result,   a stable dynamical pattern is established. The competition between the two torques becomes apparent looking into  the rotational drift velocity given by Eq.~\eqref{eq:hd_smol_rot_flux} which can be equivalently written as 
\begin{equation}
  \dot{\vec{n}} =
    \left(
      \alpha_{\text{e}} \vec{n} \times \hat{\vec{B}} 
      + \frac{1}{2} (\nabla \times \vec{u})
    \right) \times \vec{n}.
\end{equation}
  Under conditions that  both terms compensate each other, the bracket vanishes and the orientation \(\vec{n}\) does not change any more.

\bibliography{references.bib}

\end{document}